\documentclass[journal,twoside,web]{ieeecolor}
\usepackage{tmi}
\usepackage{cite}
\usepackage{amsmath,amssymb,amsfonts}
\usepackage{algorithmic}
\usepackage{graphicx}
\usepackage{textcomp}
\def\BibTeX{{\rm B\kern-.05em{\sc i\kern-.025em b}\kern-.08em
    T\kern-.1667em\lower.7ex\hbox{E}\kern-.125emX}}
\newcommand{\Tone}{T\textsubscript{1}}
\newcommand{\Ttwo}{T\textsubscript{2}}

\newcommand{\TtwoPD}{T\textsubscript{2}~$\rightarrow$~PD}

\newcommand{\PDTtwo}{PD~$\rightarrow$~T\textsubscript{2}}
\newcommand{\TtwoPDTone}{T\textsubscript{2},~PD~$\rightarrow$~T\textsubscript{1}}
\newcommand{\TonePDTtwo}{T\textsubscript{1},~PD~$\rightarrow$~T\textsubscript{2}}
\newcommand{\ToneTtwoPD}{T\textsubscript{1},~T\textsubscript{2}~$\rightarrow$~PD}

\newcommand{\TtwoFlair}{T\textsubscript{2}$\rightarrow$FLAIR}
\newcommand{\FlairTtwo}{FLAIR$\rightarrow$T\textsubscript{2}}

\newcommand{\TtwoFlairTone}{T\textsubscript{2},~FLAIR~$\rightarrow$~T\textsubscript{1}}
\newcommand{\ToneFlairTtwo}{T\textsubscript{1},~FLAIR~$\rightarrow$~T\textsubscript{2}}
\newcommand{\ToneTtwoFlair}{T\textsubscript{1},~T\textsubscript{2}~$\rightarrow$~FLAIR}

\usepackage{subfig}
\usepackage{tikz}
\usetikzlibrary{matrix,calc}

\definecolor{brightcerulean}{rgb}{0.11, 0.62, 0.74}
\newcommand*{\revhl}{\textcolor{black}}

\usepackage{amsmath}
\usepackage{framed,multirow}
\usepackage{fixltx2e}
\usepackage{amssymb}
\usepackage{latexsym}
\usepackage{url}
\usepackage{xcolor}
\usepackage{lipsum}
\usepackage{hyperref}
\hypersetup{
    colorlinks=true,
    linkcolor=blue,
    filecolor=magenta,      
    urlcolor=cyan,
    pdftitle={ResViT},
    }
\usepackage{graphics}

\usepackage{gensymb}

\definecolor{newcolor}{rgb}{.8,.349,.1}
\def\baselinestretch{0.96}
\makeatletter
\IEEEtriggercmd{\reset@font\normalfont\fontsize{7.9pt}{8.40pt}\selectfont}
\makeatother
\IEEEtriggeratref{1}

\begin{document}
\title{ResViT: Residual vision transformers for multi-modal medical image synthesis}
\author{Onat Dalmaz, Mahmut Yurt, and Tolga \c{C}ukur$^*$, \IEEEmembership{Senior Member} \vspace{-1.5cm}
\\
\thanks{This study was supported in part by a TUBITAK BIDEB scholarship awarded to O. Dalmaz, and by TUBA GEBIP 2015 and BAGEP 2017 fellowships awarded to T. \c{C}ukur  (Corresponding author: Tolga \c{C}ukur).}
\thanks{O. Dalmaz, M. Yurt, and T. \c{C}ukur are with the Department of Electrical and Electronics Engineering, and the National Magnetic Resonance Research Center (UMRAM), Bilkent University, Ankara, Turkey (e-mails: \{onat, mahmut, cukur\}@ee.bilkent.edu.tr). T. Çukur is also with the Neuroscience Program, Sabuncu Brain Research Center, Bilkent University, TR-06800 Ankara, Turkey.}
}

\maketitle
\begin{abstract}
\revhl{Generative adversarial models with convolutional neural network (CNN) backbones have recently been established as state-of-the-art in numerous medical image synthesis tasks. However, CNNs are designed to perform local processing with compact filters, and this inductive bias compromises learning of contextual features. Here, we propose a novel generative adversarial approach for medical image synthesis, ResViT, that leverages the contextual sensitivity of vision transformers along with the precision of convolution operators and realism of adversarial learning.} ResViT's generator employs a central bottleneck comprising novel aggregated residual transformer (ART) blocks that synergistically combine residual convolutional and transformer modules. \revhl{Residual connections in ART blocks promote diversity in captured representations, while a channel compression module distills task-relevant information. A weight sharing strategy is introduced among ART blocks to mitigate computational burden. A unified implementation is introduced to avoid the need to rebuild separate synthesis models for varying source-target modality configurations.} Comprehensive demonstrations are performed for synthesizing missing sequences in multi-contrast MRI, and CT images from MRI. Our results indicate superiority of ResViT against competing CNN- and transformer-based methods in terms of qualitative observations and quantitative metrics.
\end{abstract}

\begin{IEEEkeywords}
medical image synthesis, transformer, residual, vision, adversarial, generative, unified \vspace{-0.25cm}
\end{IEEEkeywords}

\bstctlcite{IEEEexample:BSTcontrol}
\section{Introduction}
Medical imaging plays a pivotal role in modern healthcare by enabling in vivo examination of pathology in the human body. In many clinical scenarios, multi-modal protocols are desirable that involve a diverse collection of images from multiple scanners (e.g., CT, MRI) \cite{pichler2008}, or multiple acquisitions from a single scanner (multi-contrast MRI) \cite{moraal2008}. Complementary information about tissue morphology, in turn, empower physicians to diagnose with higher accuracy and confidence. Unfortunately, numerous factors including uncooperative patients and excessive scan times prohibit ubiquitous multi-modal imaging \cite{thukral2015,krupa2015}. As a result, there has been ever-growing interest in synthesizing unacquired images in multi-modal protocols from the subset of available images, bypassing costs associated with additional scans \cite{iglesias2013,huo2018}.

\par
Medical image synthesis aims to predict target-modality images for a subject given source-modality images acquired under a limited scan budget \cite{farsiu2004}. This is an ill-posed inverse problem since medical images are high dimensional, target-modality data are absent during inference, and there exist nonlinear differences in tissue contrast across modalities \cite{ye2013,catana2010,lee2017,roy2013,huang2017,huang2018}. Unsurprisingly, recent adoption of deep learning methods for solving this difficult problem has enabled major performance leaps \cite{zhao2017,jog2017,hien2015,vemulapalli2015,wu2016,alexander2014,huynh2015,coupe2013}. In learning-based synthesis, network models effectively capture a prior on the joint distribution of source-target images \cite{sevetlidis2016,chartsias2018,pgan}. Earlier studies using CNNs for this purpose reported significant improvements over traditional approaches \cite{bowles2016,chartsias2018,cordier2016,sevetlidis2016,joyce2017,wei2019}. Generative adversarial networks (GANs) were later introduced that leverage an adversarial loss to increase capture of detailed tissue structure \cite{gan,beers2018,pgan,yu2018,nie2018,armanious2019,lee2019,li2019}. Further improvements were attained by leveraging enhanced architectural designs \cite{zhou2020,lan2020,yurt2021mustgan,yang2021}, and learning strategies \cite{yu2019,mmgan,wang2020}. Despite their prowess, prior learning-based synthesis models are fundamentally based on convolutional architectures that use compact filters to extract local image features \cite{pix2pix,cyclegan}. Exploiting correlations among small neighborhoods of image pixels, this inductive bias reduces the number of model parameters to facilitate learning. However, it also limits expressiveness for contextual features that reflect long-range spatial dependencies \cite{wang2018nonlocal,kodali2018}.

\revhl{Medical images contain contextual relationships across both healthy and pathological tissues. For instance, bone in the skull or CSF in the ventricles broadly distribute over spatially contiguous or segregated brain regions, resulting in dependencies among distant voxels. While pathological tissues have less regular anatomical priors, their spatial distribution (e.g., location, quantity, shape) can still show disease-specific patterns \cite{adam2014grainger}. For instance, multiple diffuse brain lesions are present in multiple sclerosis (MS) and Alzheimer’s (AD); commonly located near periventricular and juxtacortical regions in MS, and near hippocampus, entorhinal cortex and isocortex in AD \cite{ellison2012neuropathology}. Meanwhile, few lesions manifest as spatially-contiguous clumps in cancer; with lesions typically located near the cerebrum and cerebellum in gliomas, and near the skull in meningiomas \cite{ellison2012neuropathology}. Thus, the distribution of pathology also involves context regarding the position and structure of lesions with respect to healthy tissue. In principle, synthesis performance can be enhanced by priors that capture these relationships. 
Vision transformers are highly promising for this goal since attention operators that learn contextual features can improve sensitivity for long-range interactions \cite{vit}, and focus on critical image regions for improved generalization to atypical anatomy such as lesions \cite{attention_unet}. 
However, adopting vanilla transformers in tasks with pixel-level outputs is challenging due to computational burden and limited localization \cite{trans_unet}. Recent studies instead consider hybrid architectures or computation-efficient attention operators to adopt transformers in medical imaging tasks \cite{TransGAN,TransCT,SLATER,kamran2021,ganbert,ptnet}.}  

\par
Here, we propose a novel deep learning model for medical image synthesis, ResViT, \revhl{that translates between multi-modal imaging data.} ResViT combines the sensitivity of vision transformers to global context, the localization power of CNNs, and the realism of adversarial learning. ResViT's generator follows an encoder-decoder architecture with a central bottleneck to distill task-critical information. The encoder and decoder contain CNN blocks to leverage local precision of convolution operators \cite{resnet}. The bottleneck comprises novel aggregated residual transformer (ART) blocks to synergistically preserve local and global context, with a weight-sharing strategy to minimize model complexity. To improve practical utility, a unified ResViT implementation is introduced that consolidates models for numerous source-target configurations. Demonstrations are performed for synthesizing missing sequences in multi-contrast MRI, and CT from MRI. Comprehensive experiments on imaging datasets from healthy subjects and patients clearly indicate the superiority of the proposed method against competing methods. \revhl{Code to implement the ResViT model is publicly available at \href{https://github.com/icon-lab/ResViT}{https://github.com/icon-lab/ResViT}.} 

\vspace{-2mm}
\subsection*{\textbf{Contributions} }
\begin{itemize}
    \item \revhl{We introduce the first adversarial model for medical image synthesis with a transformer-based generator to translate between multi-modal imaging data.}  
    \item We introduce novel aggregated residual transformer (ART) blocks to synergistically preserve localization and context.  
    \item We introduce a weight sharing strategy among ART blocks to lower model complexity and mitigate computational burden. 
    \item We introduce a unified synthesis model that generalizes across multiple configurations of source-target modalities.

\end{itemize}

\vspace{-2mm}
\section{Related Work}
The immense success of deep learning in inverse problems has motivated its rapid adoption in medical imaging \cite{review_1,review_2}. Medical image synthesis is a particularly ill-posed problem since target images are predicted without any target-modality data \cite{nie2018}. Earlier studies in this domain have proposed local networks based on patch-level processing \cite{hien2015,li2014,torrado2016}. While local networks offer benefits over traditional approaches, they can show limited sensitivity to broader context across images \cite{sevetlidis2016}. Later studies adopted deep CNNs for image-level processing with increasing availability of large imaging databases. CNN-based synthesis has been successfully demonstrated in various applications including synthesis across MR scanners \cite{behrami2016,bahrami2016b,nie2018,zhang2019}, multi-contrast MR synthesis \cite{bowles2016,chartsias2018,cordier2016,sevetlidis2016,joyce2017,wei2019}, and CT synthesis \cite{han2017,nie2016,arabi2018,klaser2019}. Despite significant improvements they enable, CNNs trained with pixel-wise loss terms tend to suffer from undesirable loss of detailed structure \cite{pix2pix,cyclegan,pgan}. \par To improve capture of structural details, GANs \cite{gan} were proposed to learn the distribution of target modalities conditioned on source modalities \cite{conditionalgan}. Adversarial losses empower GANs to capture an improved prior for recovery of high-spatial-resolution information \cite{pix2pix,cyclegan,pgan}. In recent years, GAN-based methods were demonstrated to offer state-of-the-art performance in numerous synthesis tasks, including data augmentation as well as multi-modal synthesis \cite{sandfort2019,frida2018,pgan,lee2019}. Important applications of GAN models include CT to PET \cite{bencohen2018,santini2020}, MR to CT \cite{jin2018,ge2019,woltering2017}, unpaired cross-modality \cite{dong2019,yang2018,hiasa2018,chartsias2017}, 3T-to-7T \cite{huy2021,xiang2018}, and multi-contrast MRI synthesis \cite{armanious2019,beers2018,pgan,lan2020,lee2019,li2019,mmgan,wang2020,yang2021,yu2018,yu2019,yurt2021mustgan,zhou2020,nie2018}. 
\begin{figure*}[!t]
\vspace{-2.5ex}
 \begin{minipage}[c]{0.7\textwidth}
\centerline{\includegraphics[width=0.95\textwidth]{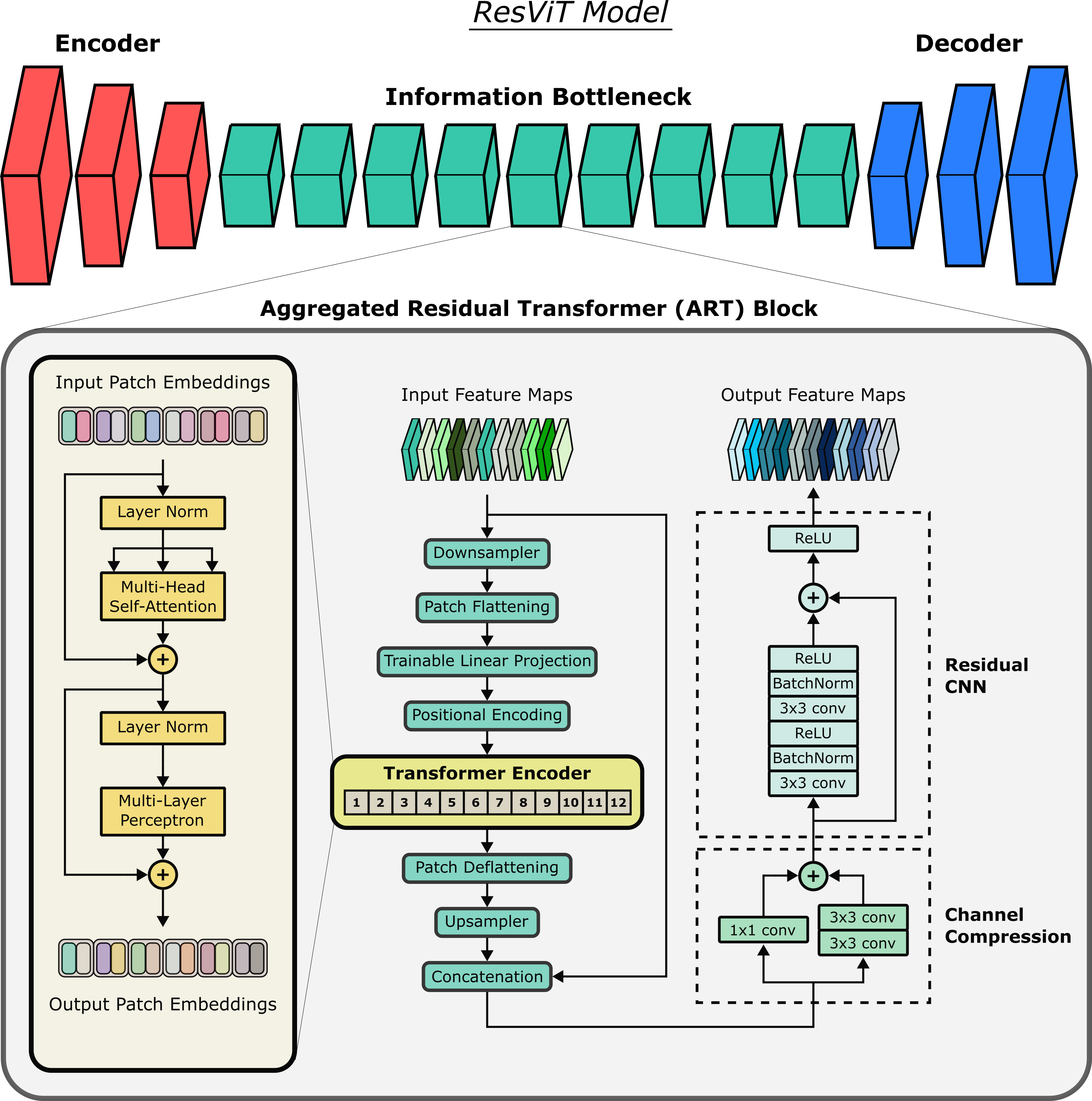}}
 \end{minipage}\hfill
 \begin{minipage}[c]{0.3\textwidth}
 \caption{The generator in ResViT follows an encoder-decoder architecture bridged with a central information bottleneck to distill task-specific information. The encoder and decoder comprise convolutional layers to maintain local precision and inductive bias in learned structural representations. Meanwhile, the information bottleneck comprises a stack of novel aggregated residual transformer (ART) blocks. ART blocks learn contextual representations via vision transformers, and synergistically fuse CNN-based local and transformer-based global representations. }
 \label{fig:main_fig}
 \end{minipage}
\vspace{-3.5ex}
\end{figure*}
\par
While GAN models have arguably emerged as a gold standard in recent years, they are not without limitation. In particular, GANs are based on purely convolutional operators known to suffer from poor across-subject generalization to atypical anatomy and sub-optimal learning of long-range spatial dependencies \cite{wang2018nonlocal,kodali2018}. Recent studies have incorporated spatial or channel attention mechanisms to modulate CNN-derived feature maps \cite{kearney2020,sagan,attention_unet,lan2020,zhao2020,yuan2020,li2020}. Such modulation motivates the network to give greater focus to regions that may suffer from greater errors \cite{sagan,attention_unet}. While attention maps might be distributed across image regions, multiplicative gating of local CNN features offers limited expressiveness in modeling of global context \cite{xie2021,trans_unet,dai2021}. 

\revhl{To incorporate contextual representations, transformer-based methods have received recent interest in imaging tasks such as segmentation \cite{trans_unet,xie2021,karimi2021}, reconstruction \cite{TransGAN,TransCT,SLATER}, and synthesis \cite{kamran2021,ganbert,ptnet}. Among relevant methods are Transformer GAN that suppresses noise in low-dose PET images \cite{TransGAN}, TransCT that suppresses noise in low-dose CT images \cite{TransCT}, and SLATER that recovers MR images from undersampled k-space acquisitions \cite{SLATER}. While these methods reconstruct images for single-modality data, ResViT translates imaging data across separate modalities. Furthermore, Transformer GAN is an adversarial model with convolutional encoder-decoder and a bottleneck that contains a transformer without external residual connections. TransCT is a non-adversarial model where CNN blocks first learn textural components of low-frequency (LF) and high-frequency (HF) image parts; and a transformer without external residual connections then combines encoded HF and textural LF maps. In comparison, ResViT is an adversarial model that employs a hybrid architecture in its bottleneck comprising a cascade of residual transformer and residual CNN modules. Unlike SLATER based on an unconditional model that maps latent variables to images via cross-attention transformers, ResViT is a conditional model based on self-attention transformers.} 

\revhl{Few recent studies have independently introduced transformer-based methods for medical image synthesis. VTGAN generates retinal angiograms from fundus photographs \cite{kamran2021} and GANBERT performs MR-to-PET synthesis \cite{ganbert}, whereas ResViT performs multi-contrast MRI and MR-to-CT synthesis. Both VTGAN and GANBERT use entirely convolutional generators and only include transformers in their discriminators. In contrast, ResViT incorporates transformers in its generator to explicitly leverage long-range context. The closest study to our work is PTNet that performs one-to-one translation between \Tone- and \Ttwo-weighted images in infant MRI \cite{ptnet}. However, PTNet is a non-adversarial model without a discriminator, and it follows a convolution-free architecture. In contrast, ResViT is an adversarial model with a hybrid CNN-transformer architecture to achieve high localization and contextual sensitivity along with a high degree of realism in synthesized images. Furthermore, a broader set of tasks are considered for ResViT including one-to-one and many-to-one translation.}

\revhl{A unique component of ResViT is the novel ART blocks in its generator that contain a cascade of transformer and CNN modules equipped with skip connections. These residual paths enable effective aggregation of contextual and convolutional representations.} Based on this powerful component, we provide the first demonstrations of a transformer architecture for many-to-one synthesis tasks and a unified synthesis model for advancing practicality over task-specific methods. 

\section{Theory and Methods}
\subsection{Residual Vision Transformers}
Here we propose a novel adversarial method for medical image synthesis named residual vision transformers, ResViT, that can unify various source-target modality configurations into a single model for improved practicality. ResViT leverages a hybrid architecture of deep convolutional operators and transformer blocks to simultaneously learn high-resolution structural and global contextual features (Fig. \ref{fig:main_fig}). \revhl{The generator subnetwork follows an encoder - information bottleneck - decoder pathway, and the discriminator subnetwork is composed of convolutional operators. The generator's bottleneck contains a stack of novel aggregated residual transformer (ART) blocks. Each ART block is organized as the cascade of a transformer module that extracts hidden contextual features, and a CNN module that extracts hidden local features of input feature maps. Importantly, external skip connections are inserted around both modules to create multiple paths of information flow through the block. These paths propagate multiple sets of features to the output: (a) Input features from the previous network layer passing through skip connections of transformer and CNN modules; (b) Contextual features computed by the transformer module passing through the skip connection of the CNN module; (c) Local features computed by the CNN module based on input features reaching through the skip connection of the transformer module; (d) Hybrid local-contextual features computed by the transformer-CNN cascade. Therefore, the main motivation for use of residual transformer and residual CNN modules in ART blocks is to learn an aggregated representation that synergistically combines lower-level input features along with their contextual, local, and hybrid local-contextual features.} 

\revhl{The central segment of ResViT containing ART blocks acts as an information bottleneck for spatial and feature dimensions of medical image representations. On the one hand, the central segment processes feature maps that have been spatially downsampled by the encoder. This increases the relative emphasis on mid- to high-level spatial information over lower-level information \cite{resnet}. On the other hand, ART blocks contain channel-compression (CC) modules that process concatenated feature maps from the previous ART block and the transformer module. CC modules downsample the concatenated maps in the feature dimension to distill a task-relevant set of convolutional and contextual features.} 

\revhl{Given the computational efficiency of convolutional layers, CNNs pervasively process feature maps at high spatial resolution to improve sensitivity for local features \cite{resnet}. In contrast, vision transformers include computationally exhaustive self-attention layers, so they typically process feature maps at relatively lower resolution \cite{vit}. To ensure that both the residual CNNs and transformers in ART blocks receive input feature maps at their expected resolutions, we incorporated down and upsampling blocks respectively at the input and output of transformer modules. This design ensures compatibility between the resolutions of feature maps extracted from CNN and transformer modules.} In the remainder of this section, we explain the detailed composition of each architectural component, and we describe the loss functions to train ResViT.  

\begin{figure*}[!t]
\vspace{-2.5ex}
\centerline{\includegraphics[width=0.9\textwidth]{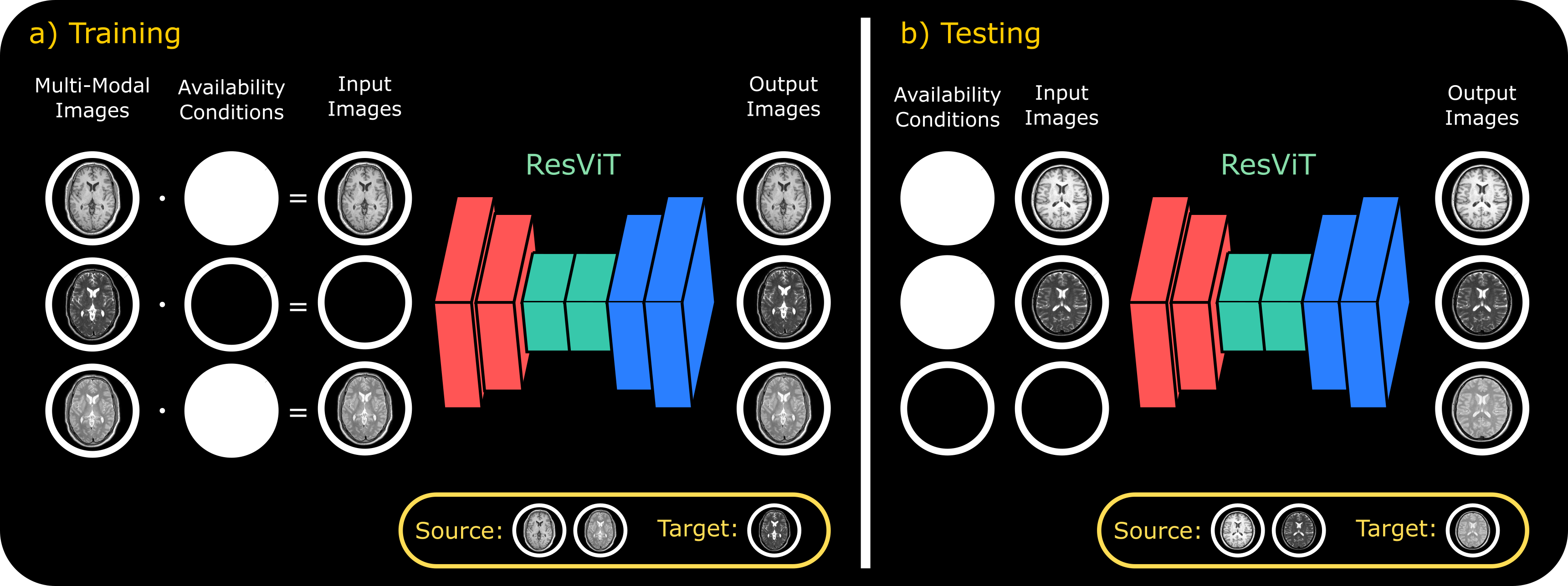}}
\caption{ResViT is a conditional image synthesis model that can unify various source-target modality configurations into a single model for improved practicality. a) During training, ResViT takes as input the entire set of images within the multi-modal protocol, including both source and target modalities. For model consolidation across multiple synthesis tasks, various configurations of source-target modalities are expressed in terms of availability conditions in ResViT. b) During inference, the specific source-target configuration is determined via the availability conditions in each given test subject.}
\label{fig:main_fig2}
\vspace{-3.5ex}
\end{figure*}

\subsubsection{Encoder} The first component of ResViT is a deep encoder network that contains a series of convolutional layers to capture a hierarchy of localized features of source images. Note that ResViT can serve as a unified synthesis model, so its encoder receives as input the full set of modalities within the imaging protocol, both source and target modalities (Fig. \ref{fig:main_fig2}). Source modalities are input via an identity mapping, whereas unavailable target modalities are masked out:
\revhl{
\begin{align}
    X_i^G = a_i \cdot m_i
\end{align}
}
where $i$ denotes the channel index of the encoder input $i\in\{1,2,\dots,I\}$, $\boldsymbol{m_i}$ is the image for the $i$th modality. In Eq. (1), $a_i$ denotes the availability of the $i$th modality:
\begin{align}
    a_i =
    \begin{cases}
      1 & \text{if $m_i$ is a source modality}\\
      0 & \text{if $m_i$ is a target modality}
    \end{cases} 
\end{align}
During training, various different configurations of source-target modalities are considered within the multi-modal protocol (e.g., \ToneTtwoPD; \TtwoPDTone; \TonePDTtwo~for a three-contrast MRI protocol). During inference, the specific source-target configuration is determined via the availability conditions in individual test subjects. Given the availability-masked multi-channel input, the encoder uses convolutional operators to learn latent structural representations shared across the consolidated synthesis tasks. \revhl{The encoder maps the multi-channel input $X^G$ onto the embedded latent feature map $f_{n_e}\in\mathbb{R}^{N_C,H,W} $ via convolutional filters,}
where $N_C$ is the number of channels, $H$ is the height and $W$ is the width of the \revhl{feature map}. These representations are then fed to the information bottleneck.

\subsubsection{Information Bottleneck} Next, ResViT employs a residual bottleneck to distill task-relevant information in the encoded features. Note that convolution operators have greater power in capturing localized features, whereas attention operators are more sensitive to context-driven features. To simultaneously maintain localization power and contextual sensitivity, we introduce ART blocks that aggregate the information from residual convolutional and transformer branches (Fig. \ref{fig:main_fig}). Receiving as input the $j$th layer feature maps $f_j\in\mathbb{R}^{N_C,H,W}$, an ART block first processes the feature maps via a vision transformer. Due to computational constraints, the transformer expects feature maps at smaller resolutions compared to convolutional layers. Thus, the spatial dimensions ($H,W$) of $f_j\in\mathbb{R}^{N_C,H,W}$ are lowered by a downsampling block ($\textit{\textbf{DS}}$):
\begin{align}
    f_j'\in\mathbb{R}^{N_C',H',W'}=\textit{\textbf{DS}}(f_j)
\end{align}
where $\textit{\textbf{DS}}$ is implemented as a stack of strided convolutional layers, $f_j'\in\mathbb{R}^{N_C',H',W'}$ are downsampled feature maps with $W'=W/M$, $H'=H/M$, $M$ denoting the downsampling factor. A transformer branch then processes $f_j'$ to extract contextual information. Accordingly, $f_j'$ is first split into \revhl{$N_P=W'H'/P^2$} non-overlapping patches of size ($P,P$), and the patches are then flattened \revhl{to $N_C' P^2$-dimensional vectors}. The transformer embeds patches onto an $N_D$-dimensional space via trainable linear projections, supplemented with learnable positional encoding: 
\begin{align}
z_0 = [f_j^1P_E;f_j^2P_E;\dots;f_j^{N_P}P_E]+P_E^{pos}
\end{align}
where $z_0\in\mathbb{R}^{N_P,N_D}$ are the input patch embeddings, \revhl{$f_j^p\in\mathbb{R}^{N_C' P^2}$ is the $p$th patch}, $P_E$ is the embedding projection, and $P_E^{pos}$ is the learnable positional encoding. 

\par Next, the transformer encoder processes patch embeddings via a cascade of $L$ layers of multi-head self-attention ($\textit{\textbf{MSA}}$) \cite{vaswani2017} and multi-layer perceptrons ($\textit{\textbf{MLP}}$) \cite{mlp}. The output of the $l$th layer in the transformer encoder is given as:
\begin{align}
&z_l' = \textit{\textbf{MSA}}(\textit{\textbf{LN}}(z_{l-1}))+z_{l-1}\\
&z_l = \textit{\textbf{MLP}}(\textit{\textbf{LN}}(z_l'))+z_l'
\end{align}
$\textit{\textbf{MSA}}$ layers in Eq. 5 employ $S$ separate self-attention heads:
\begin{align}
\textit{\textbf{MSA}}(z) = [\textit{\textbf{SA}}_1(z);\textit{\textbf{SA}}_2(z);\dots;\textit{\textbf{SA}}_S(z)]U_{msa}
\end{align}
where $\textit{\textbf{SA}}_s$ stands for the $s$th attention head with $s\in\{1,2,\dots,S\}$ and $U_{msa}$ denotes the learnable tensor projecting attention head outputs. $\textit{\textbf{SA}}$ layers compute a weighted combination of all elements of the input sequence $z$: $\textit{\textbf{SA}}(z) = Av$ where \revhl{$v$ is value, and} attention weights $A_{a,b}$ are taken as pairwise similarity between the query $q$ and key $k$:
\begin{align}
A_{a,b} = softmax(q_a\,k_b^T / {N_D}^{0.5})
\end{align}
\revhl{Note that $q$, $k$, $v$ are respectively obtained as learnable projections $T_q$, $T_k$, $T_v$ of $z$.}
\par 
The output of the transformer encoder $z_L$ is then deflattened \revhl{to form}
$g_j'\in\mathbb{R}^{N_D,H',W'}$. Resolution of $g_j'$ is increased to match the size of input feature maps via an upsampling block $\textit{\textbf{US}}$ based on transposed convolutions:
\begin{align}
g_j\in\mathbb{R}^{N_C,H,W}=\textbf{\textit{US}}(g_j')
\end{align}
where $g_j\in\mathbb{R}^{N_C,H,W}$ are upsampled feature maps output by the transformer module. \revhl{Channel-wise concatenation is performed to fuse global context learned via the transformer with localized features captured via convolutional operators.}
\revhl{To distill learned structural and contextual representations}, the channels of the concatenated feature maps are then compressed via a channel compression ($\textit{\textbf{CC}}$) module:
\revhl{\begin{align}
h_j\in\mathbb{R}^{N_C,H,W}=\textit{\textbf{CC}}(concat(f_j,g_j))
\end{align}}
where $h_j$ are compressed feature maps. $\textit{\textbf{CC}}$ uses two parallel convolutional branches of varying kernel size. Finally, the feature maps are processed via a residual CNN (\textit{\textbf{ResCNN}}) \cite{resnet}:
\begin{align}
f_{j+1}\in\mathbb{R}^{N_C,H,W}=\textit{\textbf{ResCNN}}(h_j)
\end{align}
where $f_{j+1}$ denotes the output of the ART block at the $j$th network layer. 

\subsubsection{Decoder}The last component of the generator is a deep decoder based on transposed convolutional layers. Because ResViT can serve as a unified model, its decoder can synthesize all contrasts within the multi-modal protocol regardless of the specific source-target configuration (Fig. \ref{fig:main_fig2}). The decoder receives as input the feature maps \revhl{$f_A$} distilled by the bottleneck and produces multi-modality images \revhl{$\hat{Y}^G_i\in \hat{Y}^G$} in separate channels, \revhl{where $A$ is the total number of ART blocks, and $\hat{Y}^G_i$ denotes the $i$th synthesized modality.}
\subsubsection{Parameter Sharing Transformers}
Multiple ART blocks are used in the information bottleneck to increase the capacity of ResViT in learning contextual representations. That said, multiple independent transformer blocks would inevitably elevate memory demand and risk of overfitting due to an excessive number of parameters. To prevent these risks, \revhl{a weight-sharing strategy is adopted where the model weights for the transformer encoder are tied across separate ART blocks. The tied parameters include the projection matrices $T_q$, $T_k$, $T_v$ for query, key, value along with projection tensors for attention heads $U_{msa}$ in $MSA$ layers, and weight matrices in $MLP$ layers. Remaining parameters in transformer modules including down/upsampling blocks, patch embeddings and positional encodings are kept independent. During backpropagation, updates for tied weights are computed based on the summed error gradient across ART blocks.}  

\subsubsection{Discriminator}The discriminator in ResViT is based on a conditional PatchGAN architecture \cite{pix2pix}. The discriminator performs patch-level differentiation between acquired and synthetic images. This implementation increases sensitivity to localized details related to high-spatial-frequency information. As ResViT can serve as a unified model by generating all modalities in the multi-modal protocol including sources, an availability-guided selective discriminator is employed:
\revhl{\begin{align}
&X_i^D(source) = X_i^G= a_i \cdot m_i\\
&X_i^D(syn\,target) = (1-a_i) \cdot Y_i^G\\
&X_i^D(acq\,target) = (1-a_i) \cdot m_i
\end{align}}
where $X_i^D(source)$ are source images, $X_i^D(syn\,target)$ are synthesized target images, and $X_i^D(acq\,target)$ are acquired target images. The conditional discriminator receives as input the concatenation of source and target images:
\begin{align}
&X^D(synthetic)=concat(X_i^D(source),X_i^D(syn\,target)) \\ &X^D(acquired)=concat(X_i^D(source),X_i^D(acq\,target))
\end{align}
where $X^D(synthetic)$ is the concatenation of source and synthetic target images, and $X^D(acquired)$ is the concatenation of the source and acquired target images.

\subsubsection{Loss Function} The first term in the loss function is a pixel-wise $L_1$ loss defined between the acquired and synthesized target modalities:
\begin{align}
L_{pix}=\sum_{i=1}^{I}(1-a_i)\mathrm{E}[||G(X^G)_i-m_i||_1]
\end{align}
where $\mathrm{E}$ denotes expectation, and $G$ denotes the generator subnetwork in ResViT. ResViT takes as input source modalities to reconstruct them at the output. Thus, the second term is a pixel-wise consistency loss between acquired and reconstructed source modalities based on an $L_1$ distance:
\begin{align}
L_{rec}=\sum_{i=1}^{I}a_i\mathrm{E}[||G(X^G)_i-m_i||_1]
\end{align}
The last term is an adversarial loss defined via the conditional discriminator ($D$):
\begin{align}
\begin{split}
L_{adv}=&-\mathrm{E}[D(X^D(acquired)^2]  \\ 
&-\mathrm{E}[(D(X^D(synthetic))-1)^2]
\end{split}
\end{align}
The three terms are linearly combined to form the overall objective:
\begin{align}
L_{ResViT} = \lambda_{pix}L_{pix}+\lambda_{rec}L_{rec}+\lambda_{adv}L_{adv}
\label{eq:trainloss}
\end{align}
where \revhl{$\lambda_{pix}$, $\lambda_{rec}$, and $\lambda_{adv}$ are the weightings of the pixel-wise, reconstruction, and adversarial losses, respectively.}


\vspace{-3mm}
\subsection{Datasets}
We demonstrated the proposed ResViT model on two multi-contrast brain MRI datasets (IXI: https://brain-development.org/ixi-dataset/, BRATS \cite{brats_1,brats_2,brats_3}) and a multi-modal pelvic MRI-CT dataset \cite{mr_ct_dataset}.
\subsubsection{\revhl{IXI Dataset}} \revhl{\Tone-weighted, \Ttwo-weighted, and PD-weighted brain MR images from $53$ healthy subjects were analyzed. $25$ subjects were reserved for training, $10$ were reserved for validation, and $18$ were reserved for testing. From each subject, $100$ axial cross-sections containing brain tissues were selected. Acquisition parameters were as follows. \Tone-weighted images: TE = 4.603ms, TR = 9.813ms, spatial resolution = 0.94$\times$0.94$\times$1.2mm$^3$.
\Ttwo-weighted images: TE = 100ms, TR = 8178.34ms, spatial resolution = 0.94$\times$0.94$\times$1.2mm$^3$.
PD-weighted images: TE = 8ms, TR = 8178.34ms, spatial resolution = 0.94$\times$0.94$\times$1.2mm$^3$. The multi-contrast images in this dataset were unregistered. Hence, \Ttwo- and PD-weighted images were spatially registered onto \Tone-weighted images prior to modelling. Registration was performed via an affine transformation in FSL \cite{fslcitation} based on mutual information.} 

\subsubsection{\revhl{BRATS Dataset}} \revhl{\Tone-weighted, \Ttwo-weighted, post-contrast \Ttwo-weighted, and \Ttwo~Fluid Attenuation Inversion Recovery (FLAIR) brain MR images from $55$ subjects were analyzed. $25$ subjects were reserved for training, $10$ were reserved for validation, and $20$ were reserved for testing. From each subject, $100$ axial cross-sections containing brain tissues were selected. Please note that the BRATS dataset contains images collected under various clinical protocols and scanners at multiple institutions. As publicly shared, multi-contrast images are co-registered to the same anatomical template, interpolated to 1$\times$1$\times$1mm$^3$ resolution and skull-stripped.}

\subsubsection{\revhl{MRI-CT Dataset}}
\revhl{\Ttwo-weighted MR and CT images of the male pelvis from $15$ subjects were used. $9$ subjects were reserved for training, $2$ were reserved for validation, and $4$ were reserved for testing. From each subject, $90$ axial cross-sections were analysed. Acquisition parameters were as follows. \Ttwo-weighted images: Group 1, TE = 97ms, TR = 6000-6600ms, spatial resolution = 0.875$\times$0.875$\times$2.5mm$^3$. Group 2, TE = 91-102ms, TR = 12000-16000ms, spatial resolution = 0.875-1.1$\times$0.875-1.1$\times$2.5mm$^3$. CT images: Group 1, spatial resolution = 0.98$\times$0.98$\times$3mm$^3$, Kernel = B30f. Group 2: spatial resolution = 0.1$\times$0.1$\times$2mm$^3$, Kernel = FC17. This dataset contains images collected under various protocols and scanners for each modality. As publicly shared, multi-modal images are co-registered onto \Ttwo-weighted MR scans.}

\vspace{-4mm}
\subsection{Competing Methods}
We demonstrated the proposed ResViT model against several state-of-the-art image synthesis methods. The baseline methods included convolutional models (task-specific models: pGAN \cite{pgan}, pix2pix \cite{pix2pix}, \revhl{medSynth \cite{nie2018}}; unified models: MM-GAN \cite{mmgan}, pGAN\textsubscript{uni}), attention-augmented convolutional models (A-UNet \cite{attention_unet}, SAGAN \cite{sagan}), and transformer models (task-specific: TransUNet \cite{trans_unet}, \revhl{PTNet \cite{ptnet};} unified: TransUNet\textsubscript{uni}). Hyperparameters of each competing method were optimized via identical cross-validation procedures.

\subsubsection{\revhl{Convolutional models}}
~\\ \revhl{\textbf{\quad pGAN} A convolutional GAN model with ResNet backbone was considered \cite{pgan}. pGAN comprises CNN-based generator and discriminator networks. Its generator follows an encoder-bottleneck-decoder pathway, where the encoder and decoder are identical to those in ResViT. The bottleneck contains a cascade of residual CNN blocks.} 

\revhl{\textbf{pix2pix} A convolutional GAN model with U-Net backbone was considered \cite{pix2pix}. pix2pix has a CNN-based generator with an encoder-decoder structure tied with skip connections.} 

\revhl{\textbf{medSynth} A convolutional GAN model with residual U-Net backbone was considered as provided at https://github.com/ginobilinie/medSynthesisV1 \cite{nie2018}. The generator of medSynth contains a long-skip connection from the first to the last layer.}

\revhl{\textbf{MM-GAN} A unified synthesis model based on a convolutional GAN was considered \cite{mmgan}. MM-GAN comprises CNN-based generator and discriminator networks, where the generator is based on U-Net. MM-GAN trains a single network under various source-target modality configurations. The original MM-GAN architecture was directly adopted, except for curriculum learning to ensure standard sample selection for all competing methods. The unification strategy in MM-GAN matches the unification strategy in ResViT.} 

\revhl{\textbf{pGAN\textsubscript{uni}} A unified version of the pGAN model was trained to consolidate multiple synthesis tasks. The unification procedure was identical to that of ResViT.}

\subsubsection{\revhl{Attention-augmented convolutional models}}
~\\ \revhl{\textbf{\quad Attention U-Net (A-UNet)}  A CNN-based U-Net architecture with additive attention gates was considered \cite{attention_unet}. Here we adopted the original A-UNet model as the generator of a conditional GAN model, where the discriminator was identical to that in ResViT. }

\revhl{\textbf{Self-Attention GAN (SAGAN)} A CNN-based GAN model with self-attention modules incorporated into the generator was considered \cite{sagan}. Here we adapted the original SAGAN model designed for unconditional mapping by inserting the self-attention modules into the pGAN model as described in \cite{scgan}. For fair comparison, the number and position of attention modules in SAGAN were matched to those of transformer modules in ResViT.}

\subsubsection{\revhl{Transformer models}}
~\\ \revhl{\textbf{\quad TransUNet} A recent hybrid CNN-transformer architecture was considered  \cite{trans_unet}. Here, we adopted the original TransUNet model as the generator of a conditional GAN architecture with an identical discriminator to ResViT. We further replaced the segmentation head with a convolutional layer for synthesis.}

\revhl{\textbf{PTNet} A recent convolution-free transformer architecture was considered \cite{ptnet}. Here we adopted the original PTNet model as the generator of a conditional GAN architecture with an identical discriminator to ResViT.}

\revhl{\textbf{TransUNet\textsubscript{uni}} The TransUNet model was unified to consolidate multiple synthesis tasks. The unification procedure was identical to that of ResViT.
}

\vspace{-2mm}
\subsection{Architectural Details}

\revhl{The encoder in the ResViT model contained three convolutional layers of kernel size 7, 3, 3 respectively. The feature map in the encoder output was of size $\mathbb{R}^{256,64,64}$, and this dimensionality was retained across the information bottleneck. The decoder contained three convolutional layers of kernel size 3, 3, 7 respectively. The information bottleneck contained nine ART blocks. The downsampling blocks preceding transformers contained two convolutional layers with stride 2 and kernel size 3. The upsampling blocks succeeding transformers contained two transposed convolutional layers with stride 2 and kernel size 3. Down and upsampling factors were set to $M$ = 4. Channel compression lowered the number of channels from 512 to 256. The transformer encoder was adopted by extracting the transformer component of the ImageNet-pretrained model R50+ViT-B/16 (https://github.com/google-research/vision\_transformer). The transformer encoder expected an input map of 16$\times$16 spatial resolution. Patch flattening was performed with size $P$ = 1 yielding a sequence length of 256 \cite{vit}. Note that transformer modules contain substantially higher number of parameters compared to convolutional modules. Thus, retaining a transformer in each ART block results in significant model complexity, inducing computational burden and suboptimal learning. To alleviate these issues, transformer modules in ART blocks utilized tied weights, and they were only retained in a subset of ART blocks while remaining blocks reduced to residual CNNs.} 

\renewcommand{\arraystretch}{1.20}
\begin{table}[th]
\centering
\captionsetup{justification   = justified,singlelinecheck = false}
\resizebox{0.85\columnwidth}{!}{%
\begin{tabular}{cccc}
\hline
\multirow{2}{*}{Configuration}          & \ToneTtwoPD & \ToneTtwoFlair & MRI $\rightarrow$ CT \\ \cline{2-4} 
                           & PSNR           & PSNR       & PSNR           \\ \hline
$A_1-A_5$    &   33.23        &  24.82        &   26.40              \\ \hline
$A_1-A_6$      &  \textbf{33.34}        &    \textbf{24.88}     &    26.56                          \\ \hline
$A_1-A_9$   &     33.27      &      24.77    &     \textbf{26.58}                              \\ \hline
$A_4-A_9$   &        33.11   &   24.63       &  26.19                                     \\ \hline
$A_5-A_9$    &     33.05      &   24.65       &   26.27                       \\ \hline
$A_1-A_6-A_9$  &      32.89     &   24.82       &   26.20                   \\ \hline
\end{tabular}}
\caption{\revhl{Validation performance of candidate ResViT configurations in representative synthesis tasks. Performance is taken as PSNR (dB) between synthesized and reference target images. $A_i$ denotes the presence of a transformer module in the $i$th ART block.}}
\label{tab:config_val}
\end{table}

\renewcommand{\arraystretch}{1.20}
\begin{table}[ht]
\centering
\captionsetup{justification = justified,singlelinecheck = false}
\resizebox{0.85\columnwidth}{!}{
\begin{tabular}{cccc}
\hline
\multirow{2}{*}{Transformer size}          & \ToneTtwoPD & \ToneTtwoFlair & MRI $\rightarrow CT$ \\ \cline{2-4} 
                           & PSNR           & PSNR       & PSNR           \\ \hline
Base    &   \textbf{33.34}        &  \textbf{24.88}        &   \textbf{26.56}              \\ \hline
Large      &  33.14        &    24.60     &    26.46                          \\ \hline
\end{tabular}
}
\caption{\revhl{Validation performance of ResViT models with varying sizes of transformer modules in representative synthesis tasks.}}
\label{tab:scale_val}
\vspace{-1ex}
\end{table}

\renewcommand{\tabcolsep}{3pt}
\renewcommand{\arraystretch}{1.20}
\begin{table}[ht]
\centering
\captionsetup{justification = justified,singlelinecheck = false}
\resizebox{\columnwidth}{!}{
\begin{tabular}{ccccccccc}
\hline
& ResViT & pGAN & pix2pix & medSynth & A-UNet & SAGAN & TransUNet &  PTNet\\ \hline
Time & 98  & 60  &  60 & 81 & 70  & 63  & 78 & 224   \\ \hline
\end{tabular}
}
\caption{\revhl{Average inference times (msec) of synthesis models per single cross-section.}}
\label{tab:times}
\vspace{-1ex}
\end{table}

\revhl{The configuration of transformer modules, i.e. their total number and position, was selected via cross-validation experiments. Due to the extensive number of potential configurations, a pre-selection process was implemented. Accordingly, performance for a transformer module inserted in a single ART block ($A_1,\:A_2,\:...,\: A_9$) was measured, and the top half of positions was pre-selected. Composite configurations with multiple transformer modules were then formed based on the pre-selected blocks ($A_1-A_5,\:A_1-A_6-A_9$ etc.). We observed that retaining more than 2 modules elevated complexity without any performance benefits. Validation performance for the best performing configurations ($A_1-A_5,\: A_1-A_6,\:A_1-A_9,\:A_5-A_9,\:A_4-A_9,\:A_5-A_9,\: A_1-A_6-A_9$) are listed in \revhl{Table \ref{tab:config_val}} for three representative tasks (\ToneTtwoPD~ in IXI, \ToneTtwoFlair~ in BRATS, and MRI $\rightarrow$ CT in MRI-CT). Consistently across tasks, the ($A_1-A_6$) configuration yielded near-optimal performance and so it was selected for all experiments thereafter.}

\par
\revhl{We also tuned the intrinsic complexity of transformer modules. To do this, two variant modules were examined: "base" and "large". The "base" module contained 12 layers with latent dimensionality $N_d=768$, 12 attention heads, and 3073 hidden units in each layer of the $MLP$. Meanwhile, the "large" module contained 24 layers with latent dimensionality $N_d=1024$, 16 attention heads, and 4096 hidden units in each layer of the MLP. Validation performances based on the two variant modules are listed in Table \ref{tab:scale_val}. The "base" module that offers higher performance for lower computational complexity was selected for consequent experiments.}


\vspace{-1.25ex}
\subsection{Modeling Procedures}
\revhl{For fair comparisons among competing methods,} all models were implemented adversarially using the same PatchGAN discriminator \revhl{and the loss function in Eq. \ref{eq:trainloss}.} Task-specific models used adversarial and pixel-wise losses, whereas unified models used adversarial, pixel-wise, and reconstruction losses. Learning rate, number of epochs, and loss-term weighting were selected via cross-validation. \revhl{Validation performance was measured as Peak Signal to Noise Ratio (PSNR) on three representative tasks (\ToneTtwoPD~ in IXI, \ToneTtwoFlair~ in BRATS, and MRI $\rightarrow$ CT in MRI-CT). We considered different learning rates in the set \{$10^{-5}$, $10^{-4}$, $2$x$10^{-4}$, $5$x$10^{-4}$, $10^{-3}$\} and number of epochs in the set \{5, 10, ..., 200\}. Eq. \ref{eq:trainloss} contains only two degrees of freedom regarding the loss-term weights, and prior studies have reported models with higher weighting for pixel-wise over adversarial loss \cite{pgan,yu2019}. Thus, we considered $\lambda_{pix}$ in \{20, 50, 100, 150\} and $\lambda_{adv}=1$. Note that $\lambda_{rec}=0$ by definition in task-specific models with fixed configuration of source and target modalities, while $\lambda_{rec}=\lambda_{pix}$ was used in unified models as both loss terms measure the $L_1$-norm difference between reference and generated images for individual modalities. To minimize potential biases among competing methods, a common set of parameters that consistently yielded near-optimal performance were prescribed for all methods: $2\times10^{-4}$ learning rate, $100$ training epochs, $\lambda_{adv}=1,\lambda_{pix}=100$ for task-specific models, and $\lambda_{adv}=1,\lambda_{rec}=100,\lambda_{pix}=100$ for unified models. All competing methods were trained via the Adam optimizer \cite{adam} with $\beta_1=0.5,\,\beta_2=0.999$. The learning rate was constant for the first 50 epochs and linearly decayed to $0$ in the remaining epochs. Transformer modules in TransUNet and ResViT were initiated with ImageNet pre-trained versions for object classification \cite{imagenet21k}. ART blocks were initiated without transformer modules and then fine-tuned for $50$ epochs following insertion of transformers at a higher learning rate of $10^{-3}$ as in \cite{vit}. Elevated learning rate during the second half of the training procedure was not adopted for other methods as it diminished performance.} Modelling was performed via the PyTorch framework on Nvidia RTX A4000 GPUs. Inference times are listed in Table \ref{tab:times}.

Synthesis quality was assessed via PSNR and Structural Similarity Index (SSIM) \cite{ssim}. Metrics were calculated between ground truth and synthesized target images. Mean and standard deviations of metrics were reported across an independent test set, non-overlapping with training-validation sets. Significance of performance differences were evaluated with signed-rank tests (p$<$0.05). Tests were conducted on subject-average metrics, except MRI $\rightarrow$ CT where cross-sectional metrics were tested in each subject due to limited number of test subjects.  

\vspace{-1.25ex}
\subsection{Experiments}

\subsubsection{Multi-Contrast MRI Synthesis} Experiments were conducted on the IXI and BRATS datasets to demonstrate synthesis performance in multi-modal MRI. In the IXI dataset, one-to-one tasks of \TtwoPD; \PDTtwo~and many-to-one tasks of \ToneTtwoPD; \TonePDTtwo; \TtwoPDTone~were considered. In the BRATS dataset, one-to-one tasks of \TtwoFlair; \FlairTtwo, many-to-one tasks of \ToneTtwoFlair; \ToneFlairTtwo; \TtwoFlairTone~were considered. In both datasets, task-specific ResViT models were compared against pGAN, pix2pix, \revhl{medSynth}, A-UNet, SAGAN, TransUNet, \revhl{and PTNet}. Meanwhile, unified ResViT models were demonstrated against pGAN$_{\mathrm{uni}}$, MM-GAN, and TransUNet$_{\mathrm{uni}}$. 

\begin{figure*}[t]
\vspace{-0.5mm}
\centering
\includegraphics[width=0.99\textwidth]{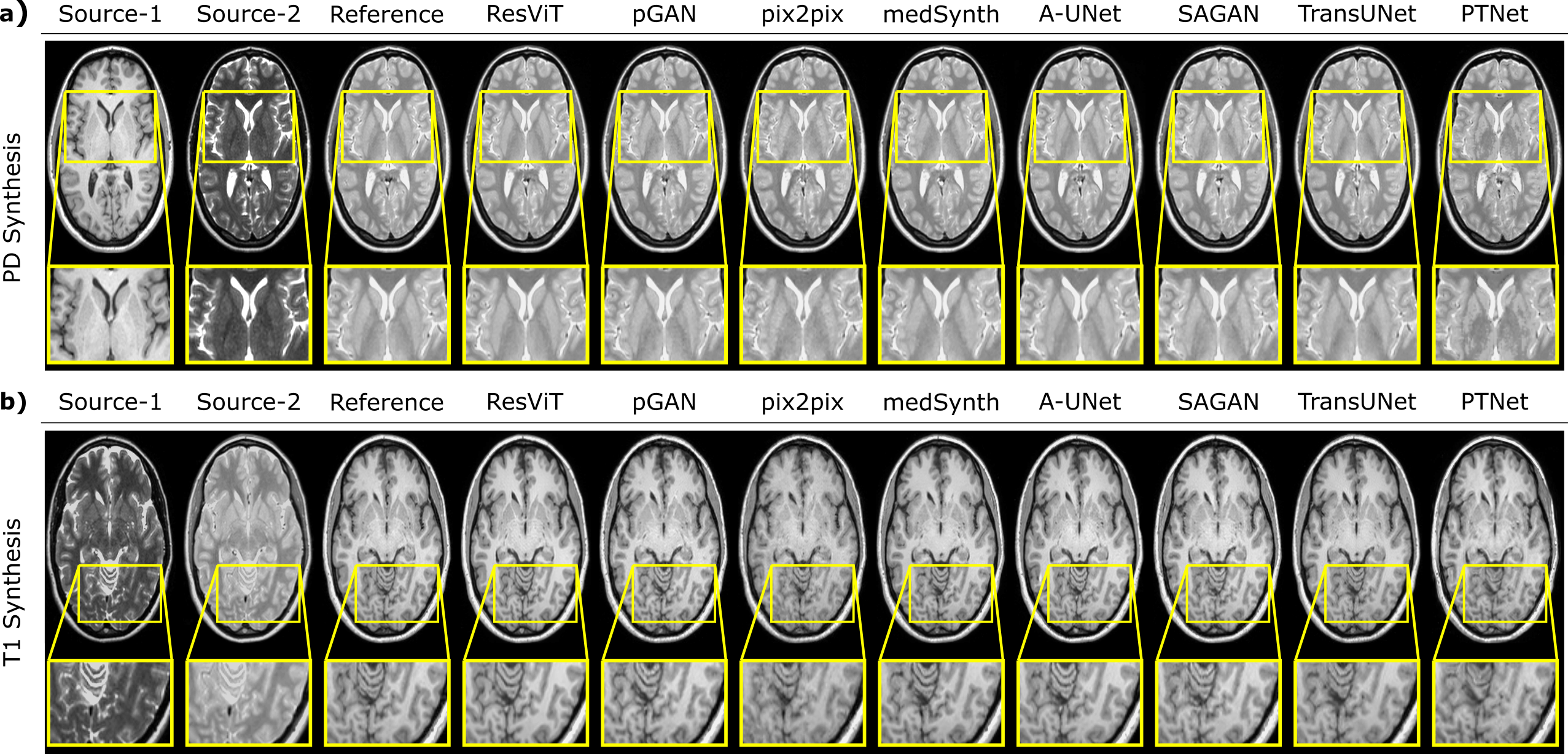}
\captionsetup{justification=justified,singlelinecheck=false}
\caption{\revhl{ResViT was demonstrated on the IXI dataset for two representative many-to-one synthesis tasks: a) \ToneTtwoPD, b) \TtwoPDTone. Synthesized images from all competing methods are shown along with the source images and the reference target image. ResViT improves synthesis performance in regions that are depicted sub-optimally in competing methods. Overall, ResViT generates images with lower artifact and noise levels and sharper tissue depiction.}}
\label{fig:IXI}
\vspace{-4mm}
\end{figure*}

\renewcommand{\tabcolsep}{2pt}
\renewcommand{\arraystretch}{1.20}
\begin{table}[]
\centering
\resizebox{1.02\columnwidth}{!}{%
\begin{tabular}{ccccccccccc}
\hline
\multirow{2}{*}{}          & \multicolumn{2}{c}{\ToneTtwoPD} & \multicolumn{2}{c}{\TonePDTtwo} & \multicolumn{2}{c}{\TtwoPDTone} & \multicolumn{2}{c}{\TtwoPD} & \multicolumn{2}{c}{\PDTtwo}  \\ \cline{2-11} 
                           & PSNR      & SSIM     & PSNR      & SSIM     & PSNR      & SSIM  & PSNR      & SSIM & PSNR      & SSIM   \\ \hline
\multirow{2}{*}{ResViT}    &   \textbf{33.92}        &  \textbf{0.977}        &   \textbf{35.71}        &    \textbf{0.977}      & \textbf{29.58 }         &  \textbf{0.952} & \textbf{32.90}         &  \textbf{0.972}  & \textbf{34.24 }         &  \textbf{0.972}     \\
                           &    \textbf{$\pm$1.44 }      &  \textbf{$\pm$0.004}        &  \textbf{$\pm$1.20}         &   \textbf{$\pm$0.005}       &   \textbf{$\pm$1.37}        &     \textbf{$\pm$0.011}  &   \textbf{$\pm$1.20}        &     \textbf{$\pm$0.005}  &   \textbf{$\pm$1.09}        &     \textbf{$\pm$0.005}     \\ \hline
\multirow{2}{*}{pGAN}      &  32.91         &    0.966      &    33.95      &    0.965      &   28.71        &    0.941 &   32.20        &    0.963 &   33.05        &    0.963     \\
                        &    $\pm$0.94       &  $\pm$0.005        &  $\pm$1.06         &   $\pm$0.006       &   $\pm$1.08        &     $\pm$0.013  &   $\pm$1.00        &     $\pm$0.005 &   $\pm$0.95        &     $\pm$0.007            \\ \hline
\multirow{2}{*}{pix2pix}   &     32.25      &      0.974    &     33.62      &    0.973      &    28.35 &     0.949 &    30.72 &     0.956
&    30.74 &     0.950\\
                           &    $\pm$1.24       &  $\pm$0.006        &  $\pm$1.31         &   $\pm$0.009       &   $\pm$1.24        &     $\pm$0.016   &   $\pm$1.28        &     $\pm$0.007&   $\pm$1.63        &     $\pm$0.012       \\ \hline
\multirow{2}{*}{medSynth} &      33.23     &   0.967       &   32.66        &  0.963        &      28.43     & 0.938   &      32.20     & 0.964 &      30.41     & 0.956       \\
                          &    $\pm$1.09       &  $\pm$0.005        &  $\pm$1.30         &   $\pm$0.007       &   $\pm$1.01        &     $\pm$0.013   &   $\pm$1.10        &     $\pm$0.006 &   $\pm$3.98        &     $\pm$0.025       \\ \hline
\multirow{2}{*}{A-UNet}    &        32.24   &   0.963       &  32.43         &      0.959    &      28.95     & 0.916 &      32.05     & 0.960
&      33.32     & 0.961\\
                           &    $\pm$0.92       &  $\pm$0.014        &  $\pm$1.36            &   $\pm$0.007       &   $\pm$1.21        &     $\pm$0.013 &   $\pm$1.04        &     $\pm$0.009 &   $\pm$1.08        &     $\pm$0.007         \\ \hline
\multirow{2}{*}{SAGAN}     &     32.50      &   0.964       &   33.71        &      0.965    &      28.62     &     0.942 &      32.07     &     0.963&      32.96     &     0.962    \\
                           &    $\pm$0.93      &  $\pm$0.005        &  $\pm$1.00         &   $\pm$0.006      &   $\pm$1.10        &     $\pm$0.013   &   $\pm$0.98        &     $\pm$0.006 &   $\pm$1.01        &     $\pm$0.007        \\ \hline
\multirow{2}{*}{TransUNet} &      32.53     &   0.968       &   32.49        &  0.960        &      28.21     & 0.941  &      30.90     & 0.960 &      31.73     & 0.958        \\
                          &    $\pm$0.97       &  $\pm$0.005        &  $\pm$1.18         &   $\pm$0.008       &   $\pm$1.30        &     $\pm$0.013   &   $\pm$1.35        &     $\pm$0.006&   $\pm$1.44        &     $\pm$0.008      \\ \hline

\multirow{2}{*}{PTNet} &      30.92     &   0.952       &   32.62        &  0.954        &      27.59     & 0.923   &      31.58     & 0.958  &      30.84     & 0.947        \\
                          &    $\pm$ 0.99       &  $\pm$0.006        &  $\pm$1.96         &   $\pm$0.019       &   $\pm$1.36        &     $\pm$0.021   &   $\pm$1.30        &     $\pm$0.007 &   $\pm$2.54        &     $\pm$0.033       \\ \hline
\end{tabular}
}
\caption{\revhl{Performance of task-specific synthesis models in many-to-one (\ToneTtwoPD, \TonePDTtwo, and \TtwoPDTone) and one-to-one (\TtwoPD~and \PDTtwo) tasks in the IXI dataset. PSNR (dB) and SSIM are listed as mean$\pm$std across test subjects. Boldface indicates the top-performing model for each task.}}
\label{tab:ixi_many}
\vspace{-2ex}
\end{table}

\subsubsection{MRI to CT Synthesis}
Experiments were performed on the MRI-CT dataset to demonstrate across-modality synthesis performance. A one-to-one synthesis task of deriving target CT images from source MR images was considered. The task-specific ResViT model was compared against pGAN, pix2pix, \revhl{medSynth}, A-UNet, SAGAN, TransUNet, \revhl{and PTNet}.

\subsubsection{Ablation Studies}
Several lines of ablation experiments were conducted to demonstrate the value of the individual components of the ResViT model, \revhl{including both architectural design elements and training strategies.} Experiments were performed on three representative tasks: namely \ToneTtwoPD~ in IXI, \ToneTtwoFlair~ in BRATS, and MRI $\rightarrow$ CT. \revhl{First, we assessed the performance contribution of the three main components in ResViT: transformer modules, convolutional modules and adversarial learning. Variant models were trained when transformer modules were ablated from ART blocks, when residual CNNs were ablated from transformer-retaining ART blocks, and when the adversarial loss term and the discriminator were ablated. In addition to PSNR and SSIM, we measured the Fréchet inception distance (FID) \cite{fid} between the synthesized and ground truth images to evaluate the importance of adversarial learning.}
\par
\revhl{Second, we probed the design and training procedures of ART blocks. We assessed the utility of tied weights across transformer modules, and multiple transformer-retaining ART blocks. Variant models were trained separately using untied weights in transformers, and based on a single transformer-retraining module at either first or sixth ART blocks. We also examined the importance of model initiation with ImageNet pre-trained transformer modules, and delayed insertion of transformer modules during training. Variant models were built by using randomly initialized transformer modules, by inserting pre-trained transformer modules into ART blocks at the beginning of training, and by inserting randomly initialized transformer modules at the beginning of training.} 
\par
\revhl{Third, we investigated the design of skip connections and down/upsampling modules. We considered benefits of external skip connections in ART blocks for residual learning. Variant models were trained by removing skip connections around either the transformer or convolution modules in ART. We also assessed alternative designs for down/upsampling modules in ART to mitigate added model complexity. In a first variant, original down/upsampling modules were replaced with unlearned maxpooling modules for downsampling and bilinear interpolation modules for upsampling. In a second variant, additional downsampling layers in the encoder and upsampling layers in the decoder were included in order to remove down/upsampling modules in ART blocks.}
  \par

\begin{figure*}[t!]
\vspace{-1ex}
\includegraphics[width=0.99\textwidth]{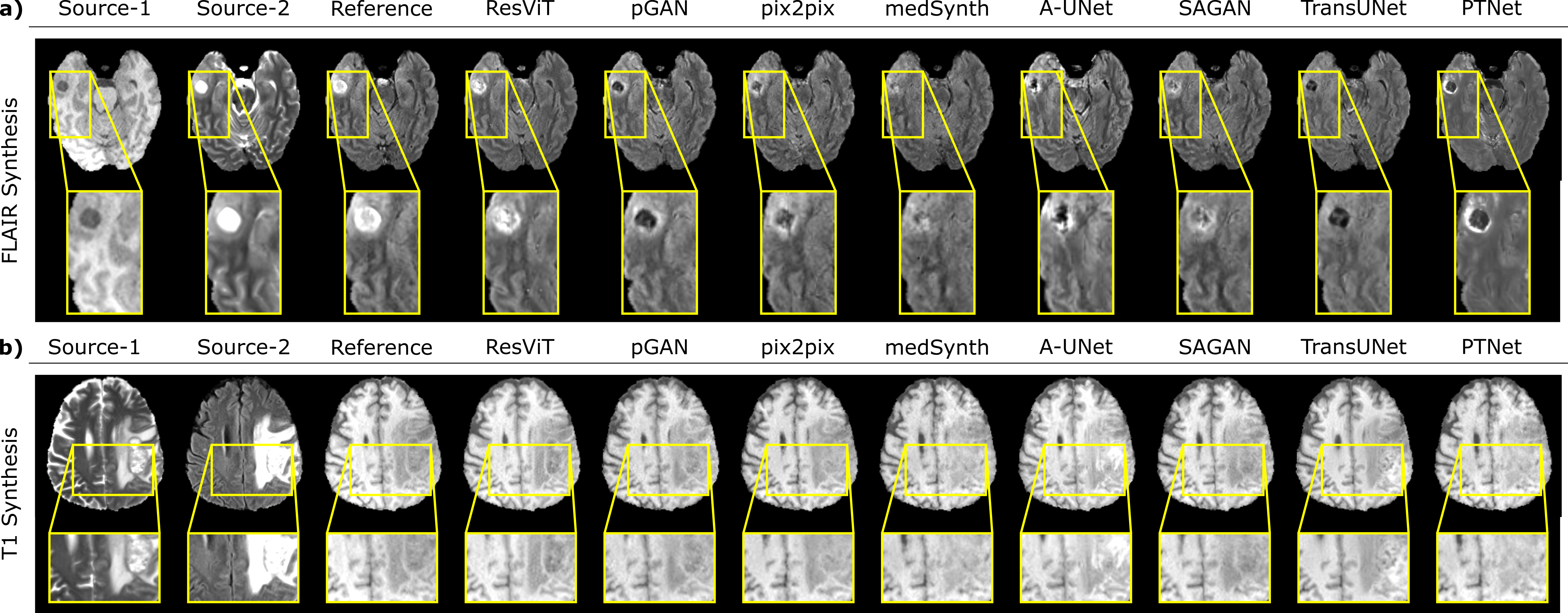}
\caption{ResViT was demonstrated on the BRATS dataset for two representative many-to-one synthesis tasks: a) \ToneTtwoFlair, b) \TtwoFlairTone. Synthesized images from all competing methods are shown along with the source images and the reference image. ResViT improves synthesis performance, especially in pathological regions (e.g., tumors, lesions) in comparison to competing methods. Overall, ResViT images have better-delineated tissue boundaries and lower artifact/noise levels.}
\label{fig:brats_many}
\vspace{-3.5ex}
\end{figure*}

 \revhl{Next, we inspected the relative strength of contextual features in the distilled task-relevant representations in ART blocks. For a quantitative assessment, we compared the $L_2$-norm of the contextual feature map derived by the transformer module against that of the input feature map to the ART block relayed through the transformer's skip connection. Note that these two maps are distilled via the channel compression (CC) module following concatenation. Thus, we also compared the $L_2$-norm of the combination weights in the CC module for the contextual versus input features.}
\par
\revhl{To interpret the information that self-attention mechanisms focus on during synthesis tasks, we computed and visualized the attention maps as captured by the transformer modules in ResViT. Attention maps were calculated based on the Attention Rollout technique, and a single average map was extracted for a given transformer module \cite{rollout}.}

\section{Results}
\subsection{Multi-Contrast MRI Synthesis}

\renewcommand{\tabcolsep}{2pt}
\renewcommand{\arraystretch}{1.20}
\begin{table}[]
\centering
\resizebox{1.02\columnwidth}{!}{%
\begin{tabular}{ccccccccccc}
\hline
\multirow{2}{*}{}          & \multicolumn{2}{c}{\ToneTtwoFlair}  & \multicolumn{2}{c}{\ToneFlairTtwo} & \multicolumn{2}{c}{\TtwoFlairTone} & \multicolumn{2}{c}{\TtwoFlair} & \multicolumn{2}{c}{\FlairTtwo} \\ \cline{2-11} 
                           & PSNR      & SSIM     & PSNR      & SSIM     & PSNR      & SSIM  & PSNR      & SSIM& PSNR      & SSIM   \\ \hline
\multirow{2}{*}{ResViT}    &   \textbf{25.84}        &  \textbf{0.886}        &   \textbf{26.90}        &    \textbf{0.938}      & \textbf{26.20}          &  \textbf{0.924}  & \textbf{24.97}          &  0.870 & \textbf{25.78}          &  \textbf{0.908}      \\
                           &    \textbf{$\pm$1.13}       &   \textbf{$\pm$0.014}        &  \textbf{$\pm$1.20 }        &   \textbf{$\pm$0.011 }      &   \textbf{$\pm$1.31 }       &     \textbf{$\pm$0.009} &   \textbf{$\pm$1.07 }       &     $\pm$0.014 &   \textbf{$\pm$0.92 }       &     \textbf{$\pm$0.015}     \\ \hline
\multirow{2}{*}{pGAN}      &  24.89         &    0.867      &    26.51      &    0.922      &   25.72        &    0.918 &   24.01        &    0.864 &   25.09        &    0.894      \\
                        &    $\pm$1.10       &  $\pm$0.015        &  $\pm$1.13         &   $\pm$0.012       &   $\pm$1.54       &     $\pm$0.011     &   $\pm$1.15       &     $\pm$0.014 &   $\pm$1.52       &     $\pm$0.015        \\ \hline
\multirow{2}{*}{pix2pix}   &     24.31      &      0.862    &     26.12      &    0.920      &    25.80      &0.918 &    23.15      &0.869&    24.52      &0.883          \\
                           &    $\pm$1.21       &  $\pm$0.015        &  $\pm$1.53         &   $\pm$0.012       &   $\pm$1.72        &     $\pm$0.011  &   $\pm$1.93        &     $\pm$0.016 &   $\pm$0.88        &     $\pm$0.014         \\ \hline
\multirow{2}{*}{medSynth} &      23.93     &   0.863       &   26.44        &  0.921        &      25.72     & 0.914   
&      23.36     & 0.864&      24.41     & 0.888\\
                          &    $\pm$1.45       &  $\pm$0.016        &  $\pm$0.76         &   $\pm$0.011      &   $\pm$1.62       &     $\pm$0.012      &   $\pm$1.88       &     $\pm$0.017 &   $\pm$0.82       &     $\pm$0.014    \\ \hline
\multirow{2}{*}{A-UNet}    &        24.36   &   0.857       &  26.48         &      0.924    &      25.67     & 0.918  &      23.69     & \textbf{0.873} &      24.56     & 0.891         \\
                           &    $\pm$1.24       &  $\pm$0.017        &  $\pm$1.21         &   $\pm$0.012       &   $\pm$1.35        &     $\pm$0.010 &   $\pm$1.57        &     \textbf{$\pm$0.015} &   $\pm$0.94        &     $\pm$0.014         \\ \hline
\multirow{2}{*}{SAGAN}     &     24.62      &   0.869       &   26.41        &      0.919    &      25.91     &     0.918  &      24.02     &     0.860 &      25.10     &     0.893    \\
                           &    $\pm$1.17       &  $\pm$0.014        &  $\pm$1.22         &   $\pm$0.012       &   $\pm$1.42       &     $\pm$0.011    &   $\pm$1.35       &     $\pm$0.015 &   $\pm$0.88       &     $\pm$0.014       \\ \hline
\multirow{2}{*}{TransUNet} &      24.34     &   0.872       &   26.51        &  0.920        &      25.76     & 0.921  &      23.70     & 0.864&      24.62     & 0.891       \\
                          &    $\pm$1.26       &  $\pm$0.014        &  $\pm$0.92         &   $\pm$0.010      &   $\pm$1.69       &     $\pm$0.011 &   $\pm$1.75       &     $\pm$0.015 &   $\pm$0.81       &     $\pm$0.015          \\ \hline
\multirow{2}{*}{PTNet} &      23.78     &   0.851       &   25.09        &  0.905        &      22.19     & 0.920  
&      23.01     & 0.851&      24.78     & 0.894       \\
                          &    $\pm$1.24       &  $\pm$0.031        &  $\pm$1.23         &   $\pm$0.016      &   $\pm$1.88       &     $\pm$0.014 &   $\pm$0.85       &     $\pm$0.014 &   $\pm$0.88       &     $\pm$0.015         \\ \hline                          
\end{tabular}}
\caption{
\revhl{Performance of task-specific synthesis models in many-to-one tasks (\ToneTtwoFlair, \ToneFlairTtwo, and \TtwoFlairTone) and one-to-one tasks (\TtwoFlair~and \FlairTtwo) across test subjects in the BRATS dataset. Boldface indicates the top-performing model for each task.}}
\label{tab:brats_many}
\vspace{-1ex}
\end{table}

\subsubsection{Task-Specific Synthesis Models}
We demonstrated the performance of ResViT in learning task-specific synthesis models for multi-contrast MRI. ResViT was compared against convolutional models (pGAN, pix2pix, \revhl{medSynth}), attention-augmented CNNs (A-UNet, SAGAN), and recent transformer architectures (TransUNet, \revhl{PTNet}). First, brain images of healthy subjects in the IXI dataset were considered. PSNR and SSIM metrics are listed in Table \ref{tab:ixi_many} for many-to-one and one-to-one tasks. ResViT achieves the highest performance in both many-to-one  (p$<$0.05) and one-to-one tasks (p$<$0.05). On average, ResViT outperforms convolutional models by \revhl{1.71}dB PSNR and \revhl{1.08}$\%$ SSIM, attention-augmented models by 1.40dB PSNR and 1.45$\%$ SSIM, and transformer models by \revhl{2.33}dB PSNR and \revhl{1.79}$\%$ SSIM (p$<$0.05). Representative images for \ToneTtwoPD~and \TtwoPDTone~are displayed in Fig. \ref{fig:IXI}a,b. Compared to baselines, ResViT synthesizes target images with lower artifact levels and sharper tissue depiction.

We then demonstrated task-specific ResViT models on the BRATS dataset containing images of glioma patients. PSNR and SSIM metrics are listed in Table \ref{tab:brats_many} for many-to-one and one-to-one tasks. ResViT again achieves the highest performance in many-to-one (p$<$0.05) and one-to-one tasks (p$<$0.05), except \TtwoFlair~where A-UNet has slightly higher SSIM. On average, ResViT outperforms convolutional models by \revhl{1.01}dB PSNR and \revhl{1.41}$\%$ SSIM, attention-augmented models by 0.84dB PSNR and 1.24$\%$ SSIM, and \revhl{transformer models} by \revhl{1.56}dB PSNR and \revhl{1.63}$\%$ SSIM (p$<$0.05). Note that the BRATS dataset contains pathology with large across-subject variability. As expected, attention-augmented models show relative benefits against pure convolutional models, yet ResViT that explicitly models contextual relationships still outperforms all baselines. Representative target images for \ToneTtwoFlair~and \TtwoFlairTone~are displayed in Fig. \ref{fig:brats_many}a,b, respectively. Compared to baselines, ResViT synthesizes target images with lower artifact levels and sharper tissue depiction. Importantly, ResViT reliably captures brain lesions in patients in contrast to competing methods with inaccurate depictions including TransUNet.

\begin{figure*}[t]
 \vspace{-0.5ex}
 \begin{minipage}[l]{0.72\textwidth}
\centerline{\includegraphics[width=0.98\textwidth]{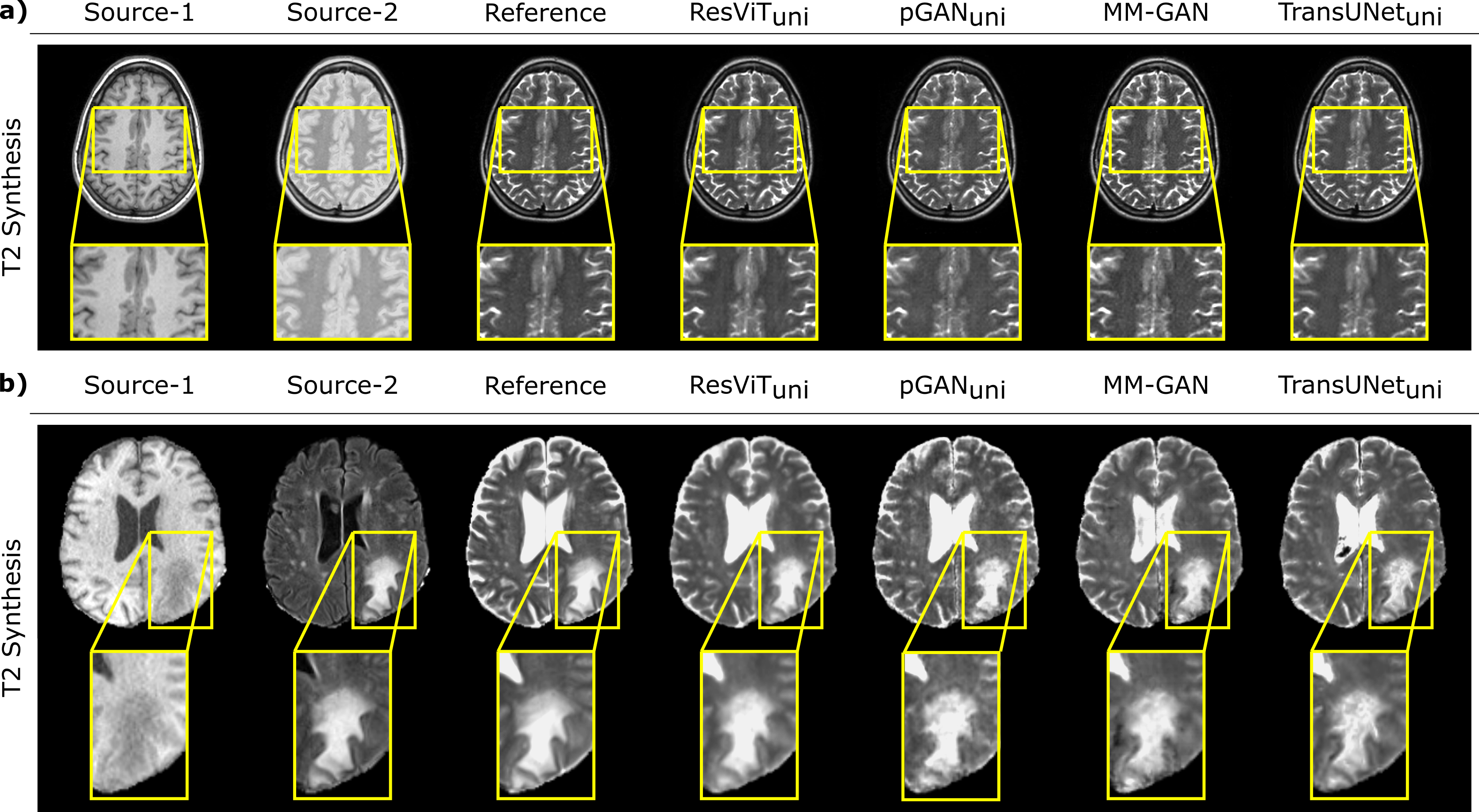}}
 \end{minipage}\hfill
 \begin{minipage}[c]{0.28\textwidth}
\caption{\revhl{ResViT$_\mathrm{uni}$ was demonstrated against other unified models on brain MRI datasets for two representative tasks: a) \TonePDTtwo~in IXI, b) \ToneFlairTtwo~in BRATS. Synthesized images from all competing methods are shown along with the source images and the reference target image. ResViT$_\mathrm{uni}$ improves synthesis performance especially in pathological regions (tumors, lesions) in comparison to competing methods. Overall, ResViT$_\mathrm{uni}$ generates images with lower artifact and noise levels and more accurate tissue depiction for tasks in both datasets.}}
\label{fig:unified}
 \end{minipage}
 \vspace{-2ex}
\end{figure*}
\par
\revhl{Superior depiction of pathology in ResViT signals the importance of ART blocks in simultaneously maintaining local precision and contextual consistency in medical image synthesis. In comparison, transformer-based TransUNet and PTNet yield relatively limited synthesis quality that might be attributed to several fundamental differences between the models. TransUNet uses only a transformer in its bottleneck while propagating shallow convolutional features via encoder-decoder skip connections, and its decoder increases spatial resolution via bilinear upsampling that might be ineffective in suppressing high-frequency artifacts \cite{upconv}. In contrast, ResViT continues encoding and propagating convolutional features across the information bottleneck to create a deeper feature representation, and it employs transposed convolutions within upsampling modules to mitigate potential artifacts. PTNet is a convolution-free architecture that relies solely on self-attention operators that have limited localization ability \cite{vit}. Instead, ResViT is devised as a hybrid CNN-transformer architecture to improve sensitivity for both local and contextual features.}

\renewcommand{\tabcolsep}{5pt}
\renewcommand{\arraystretch}{1.20}
\begin{table}[]
\centering
\resizebox{0.9\columnwidth}{!}{%
\begin{tabular}{ccccccc}
\hline
\multirow{2}{*}{}          & \multicolumn{2}{c}{\ToneTtwoPD } & \multicolumn{2}{c}{\TonePDTtwo} & \multicolumn{2}{c}{\TtwoPDTone} \\ \cline{2-7} 
                           & PSNR      & SSIM     & PSNR      & SSIM     & PSNR      & SSIM     \\ \hline
\multirow{2}{*}{ResViT$_{\mathrm{uni}}$}    &   \textbf{33.22 }       & \textbf{ 0.971 }       &  \textbf{ 33.97 }       &   \textbf{ 0.968}      & \textbf{28.80}          &  \textbf{0.946 }       \\
                           &   \textbf{ $\pm$1.21}       &  \textbf{$\pm$0.005 }       &  \textbf{$\pm$1.04}         &   \textbf{$\pm$0.006}       &   \textbf{$\pm$1.20 }       &    \textbf{ $\pm$0.013}     \\ \hline
\multirow{2}{*}{pGAN$_{\mathrm{uni}}$}     &  31.86         &    0.965      &    32.90      &    0.962      &   27.86        &    0.937      \\
                        &    $\pm$1.09      &  $\pm$0.005        &  $\pm$0.91         &   $\pm$0.006       &   $\pm$1.04        &     $\pm$0.014             \\ \hline
\multirow{2}{*}{MM-GAN }   &     30.73      &      0.955    &     30.91      &    0.951      &    27.23 &     0.925           \\
                           &    $\pm$1.16       &  $\pm$0.006        &  $\pm$1.61         &   $\pm$0.013       &   $\pm$1.24        &     $\pm$0.015         \\ \hline
\multirow{2}{*}{{TransUNet$_{\mathrm{uni}}$}} &      30.30     &   0.956       &   30.77        &  0.949        &      26.86     & 0.930         \\
                          &    $\pm$1.44       &  $\pm$0.007        &  $\pm$1.10         &   $\pm$0.014       &   $\pm$1.16        &     $\pm$0.013          \\ \hline
\end{tabular}
}
\caption{Performance of unified synthesis models in many-to-one tasks \ToneTtwoPD, \TonePDTtwo, and \TtwoPDTone) across test subjects in the IXI dataset. Boldface indicates the top-performing model for each task.
}
\label{ixi_unified}
\vspace{-1ex}
\end{table}

\subsubsection{Unified Synthesis Models}
Task-specific models are trained and tested to perform a single synthesis task to improve performance, but a separate model has to be built for each task. Next, we demonstrated ResViT in learning unified synthesis models for multi-contrast MRI. A unified ResViT (ResViT$_{\mathrm{uni}}$) was compared against unified convolutional (pGAN$_{\mathrm{uni}}$, MM-GAN) and transformer models (TransUNet$_{\mathrm{uni}}$). Performance of unified models were evaluated at test time on many-to-one tasks in IXI (Table \ref{ixi_unified}) and BRATS (Table \ref{brats_unified}). ResViT$_{\mathrm{uni}}$ maintains the highest performance in many-to-one tasks in both IXI (p$<$0.05) and BRATS (p$<$0.05). In IXI, ResViT$_{\mathrm{uni}}$ outperforms pGAN$_{\mathrm{uni}}$ by 1.12dB PSNR and 0.70$\%$ SSIM, MM-GAN by 2.37dB PSNR and 1.80$\%$ SSIM, and TransUNet$_{\mathrm{uni}}$ by 2.69dB PSNR and 1.67$\%$ SSIM (p$<$0.05). In BRATS, ResViT outperforms pGAN$_{\mathrm{uni}}$ by 0.74dB PSNR and 0.93$\%$ SSIM,  MM-GAN by 0.77dB PSNR and 0.90$\%$ SSIM, and TransUNet$_{\mathrm{uni}}$ by 1.08dB PSNR and 1.43$\%$ SSIM (p$<$0.05). Representative target images are displayed in Fig. \ref{fig:unified}. ResViT synthesizes target images with lower artifacts and sharper depiction than baselines. These results suggest that a unified ResViT model can successfully consolidate models for varying source-target configurations.

\renewcommand{\tabcolsep}{5pt}
\renewcommand{\arraystretch}{1.20}
\begin{table}[]
\centering
\resizebox{0.9\columnwidth}{!}{%
\begin{tabular}{ccccccc}
\hline
\multirow{2}{*}{}          & \multicolumn{2}{c}{\ToneTtwoFlair} & \multicolumn{2}{c}{\ToneFlairTtwo} & \multicolumn{2}{c}{\TtwoFlairTone} \\ \cline{2-7} 
                           & PSNR      & SSIM     & PSNR      & SSIM     & PSNR      & SSIM     \\ \hline
\multirow{2}{*}{ResViT$_{\mathrm{uni}}$}    &   \textbf{25.32 }       & \textbf{ 0.876}        &   \textbf{26.81 }       &    \textbf{0.921}      & \textbf{26.24 }         &  \textbf{0.922}        \\
                           &    \textbf{$\pm$0.91 }      &  \textbf{$\pm$0.015 }       & \textbf{ $\pm$1.04 }        &  \textbf{ $\pm$0.012}       &  \textbf{ $\pm$1.65}        &    \textbf{ $\pm$0.010}     \\ \hline
\multirow{2}{*}{pGAN$_{\mathrm{uni}}$}      &  24.46         &    0.865      &    26.23      &    0.914      &   25.46        &    0.912      \\
                        &    $\pm$0.99       &  $\pm$0.014        &  $\pm$1.08         &   $\pm$0.012       &   $\pm$1.20        &     $\pm$0.009             \\ \hline
\multirow{2}{*}{MM-GAN}   &     24.20      &      0.861    &     26.10      &    0.915      &    25.75      &0.916           \\
                           &    $\pm$1.34       &  $\pm$0.015      &  $\pm$1.48         &   $\pm$0.014       &   $\pm$1.64        &     $\pm$0.011          \\ \hline
\multirow{2}{*}{TransUNet$_{\mathrm{uni}}$} &      24.11     &   0.863       &   26.05        &  0.912        &      24.96     & 0.901         \\
                          &    $\pm$1.19       &  $\pm$0.014        &  $\pm$1.46         &   $\pm$0.013       &   $\pm$1.24        &     $\pm$0.012          \\ \hline
\end{tabular}
}
\caption{Performance of unified synthesis models in many-to-one tasks (\ToneTtwoFlair, \ToneFlairTtwo, and \TtwoFlairTone) across test subjects in the BRATS dataset. Boldface indicates the top-performing model for each task.}
\label{brats_unified}
\end{table}

\begin{figure*}[t!]
\vspace{-1ex}
\centerline{\includegraphics[width=1.01\textwidth]{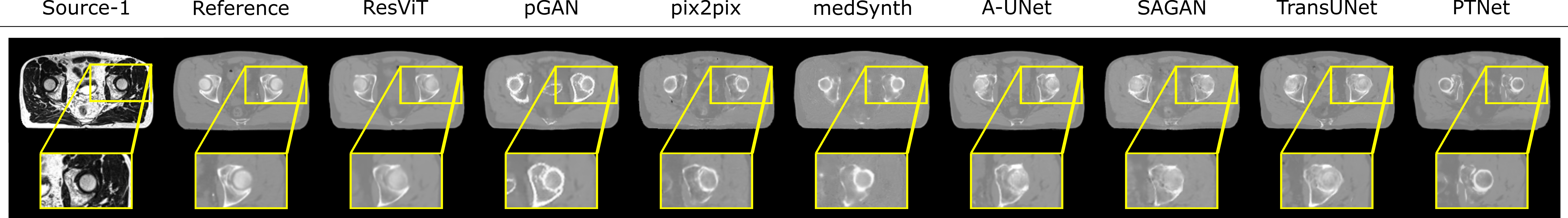}}
\caption{ResViT was demonstrated on the pelvic MRI-CT dataset for the T$_2$-weighted MRI $\rightarrow$ CT task. Synthesized images from all competing methods are shown along with the source and reference images. ResViT enhances synthesis of relevant morphology in the CT domain as evidenced by the elevated accuracy near bone structures.}
\label{fig:mr_ct}
\vspace{-3.5ex}
\end{figure*}

\renewcommand{\tabcolsep}{3pt}
\renewcommand{\arraystretch}{1.20}
\begin{table}[b]
\vspace{2mm}
\centering
\resizebox{\columnwidth}{!}{%
\begin{tabular}{cccccccccc}\hline
                                        &                       & ResViT & pGAN & pix2pix & medSynth& A-UNet & SAGAN & TransUNet  & PTNet \\ \hline
\multirow{4}{*}{\rotatebox[origin=c]{90}{MRI $\rightarrow$ CT}}\hspace{2mm} & \multirow{2}{*}{\rotatebox[origin=c]{90}{PSNR}} & \textbf{28.45}  & 26.80  &  26.53 &  26.36 & 27.80  & 27.61  & 27.76     &   26.11 \\
                                        &         & \textbf{$\pm$1.35}  & $\pm$0.90   &  $\pm$0.45 & $\pm$0.63 & $\pm$0.63   & $\pm$1.02    & $\pm$1.03   & $\pm$0.93   \\ \cline{2-10} 
                                        & \multirow{2}{*}{\rotatebox[origin=c]{90}{SSIM}} & \textbf{0.931}  & 0.905    &  0.898  & 0.894     & 0.913      & 0.910     &    0.914 & 0.900      \\
                                        &                       & \textbf{$\pm$0.009} & $\pm$0.008    & $\pm$0.004  & $\pm$0.009      & $\pm$0.004      & $\pm$0.006      & $\pm$0.009      & $\pm$0.015     \\ \hline
\end{tabular}}
\caption{Performance for the across-modality synthesis task (T$_2$-weighted MRI $\rightarrow$ CT) across test subjects in the pelvic MRI-CT dataset. Boldface indicates the top-performing model for each task.}
\label{tab:mr_ct}
\vspace{-3mm}
\end{table}

\vspace{-1ex}
\subsection{Across-Modality Synthesis}
We also demonstrated ResViT in across-modality synthesis. T$_2$-weighted MRI and CT images in the pelvic dataset were considered. ResViT was compared against pGAN, pix2pix, \revhl{medSynth}, A-UNet, SAGAN, TransUNet, \revhl{and PTNet}. PSNR and SSIM metrics are listed in Table \ref{tab:mr_ct}. ResViT yields the highest performance in each subject (p$<$0.05). On average, ResViT outperforms convolutional models by \revhl{1.89}dB PSNR and \revhl{3.20}$\%$ SSIM,  attention-augmented models by 0.75dB PSNR and 1.95$\%$ SSIM, and \revhl{transformer models} by \revhl{1.52}dB PSNR and \revhl{2.40}$\%$ SSIM (p$<$0.05). Representative target images are displayed in Fig. \ref{fig:mr_ct}. Compared to baselines, ResViT synthesizes target images with lower artifacts and more accurate tissue depiction.
Differently from multi-contrast MRI, attention-augmented models and TransUNet offer more noticeable performance benefits over convolutional models. That said, ResViT still maintains further elevated performance, particularly near bone structures in CT images. This finding suggests that the relative importance of contextual representations is higher in MRI-CT synthesis. With the help of its residual transformer blocks, ResViT offers reliable performance with accurate tissue depiction in this task.

\vspace{-1mm}
\subsection{Ablation Studies}
\begin{figure*}[t]
\centering
\includegraphics[width=\textwidth]{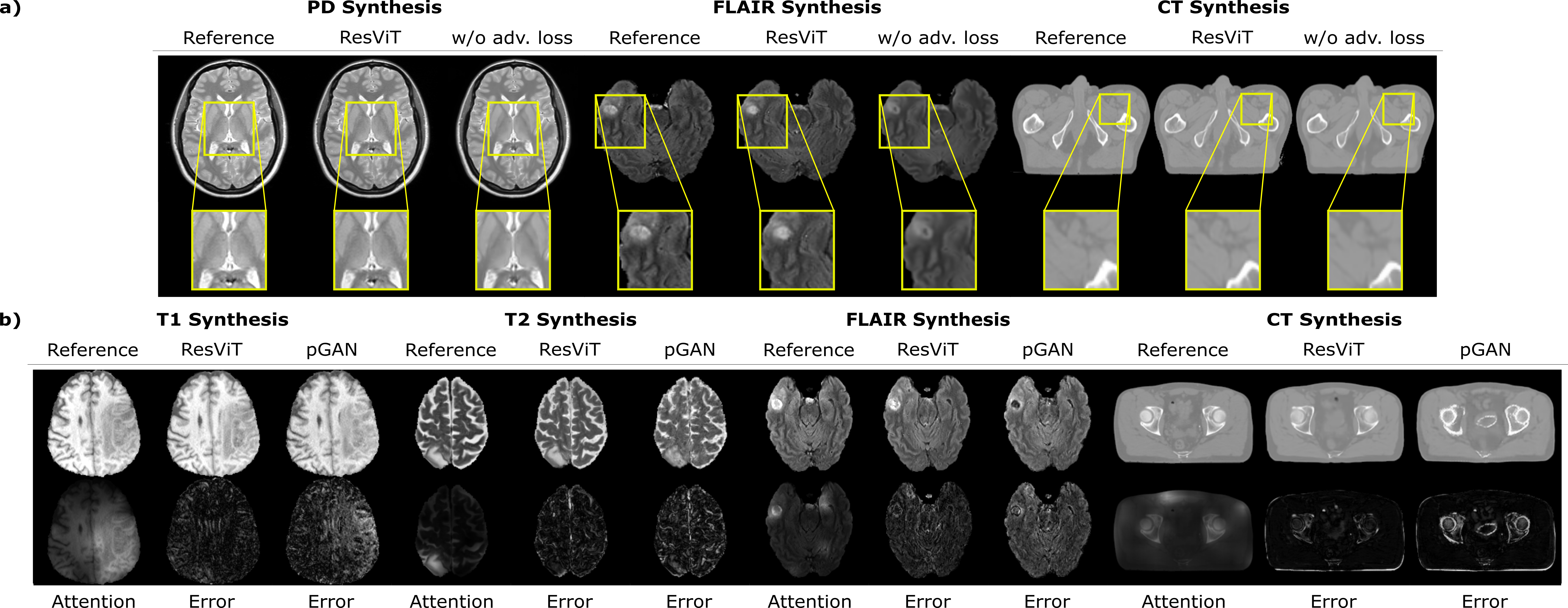}
\captionsetup{justification=justified,singlelinecheck=false}
\caption{\revhl{a) ResViT was compared against a variant where the adversarial term was removed from the loss function. Representative results are shown for \ToneTtwoPD ~in IXI, \ToneTtwoFlair ~in BRATS, and MRI $\rightarrow$ CT in the pelvic dataset. Adversarial loss improves the acuity of synthesized images. b) Representative results from ResViT and pGAN are shown along with the reference images for \TtwoFlairTone, \ToneFlairTtwo, and \ToneTtwoFlair ~in BRATS; and MRI $\rightarrow$ CT in the pelvic dataset. Error maps between the synthetic and reference images for each method are displayed, along with the attention map for the first transformer module of ResViT. Here, the attention maps were overlaid onto the reference image for improved visualization. Attention maps focus on image regions where ResViT substantially reduces synthesis errors compared to pGAN.
}}
\label{fig:adv_attention_merge}
\vspace{-11mm}
\end{figure*}

We performed a systematic set of experiments to demonstrate the added value of the main components and training strategies used in ResViT. First, we compared ResViT against ablated variants where the convolutional modules in ART blocks, transformer modules in ART blocks, or the adversarial term in training loss were separately removed. Table \ref{tab:ablation_1} lists performance metrics in the test set for three representative synthesis tasks. Consistently across tasks, ResViT yields optimal or near-optimal performance. ResViT achieves higher PSNR and SSIM in representative tasks compared to variants without transformer or convolutional modules (p$<$0.05). It also yields lower FID than these variants, except in MRI $\rightarrow$ CT where ablation of the convolutional module slightly decreases FID. Importantly, ResViT maintains notably lower FID compared to the variant without adversarial loss (albeit slightly lower SSIM in \ToneTtwoFlair~ and PSNR, \revhl{SSIM} in MRI $\rightarrow$ CT). This is expected since FID is generally considered as a more suited metric to examine the perceptual benefits of adversarial learning than PSNR or SSIM that reflect heavier influence from relatively lower frequencies \cite{fid}. Representative synthesized images are also displayed in Fig. \ref{fig:adv_attention_merge}a. ResViT images more closely mimic the reference images, and show greater spatial acuity compared against the variant without adversarial loss. Taken together, these results indicate that adversarial learning enables ResViT to more closely capture the distributional properties of target-modality images.

Second, we compared ResViT against ablated variants where the weight tying procedure across transformer modules was neglected, or transformer modules in one of the two retaining ART blocks were removed. Table \ref{tab:ablation_2} lists performance metrics in the test set. \revhl{ResViT yields higher performance than variants across representative tasks (p$<$0.05), except for the variant only retraining $A_6$ that yields similar SSIM in \ToneTtwoPD}. These results demonstrate the added value of the weight tying procedure and the transformer configuration in ResViT. We also compared ResViT against ablated variants where the pre-training of transformer modules or their delayed insertion during training were selectively neglected, as listed in Table \ref{tab:ablation_3}. \revhl{Our results indicate that ResViT outperforms all ablated variants (p$<$0.05), except for the variant without delayed insertion that yields similar SSIM in \ToneTtwoPD}.

\renewcommand{\tabcolsep}{4pt}
\renewcommand{\arraystretch}{1.20}
\begin{table}[th]
\centering
\captionsetup{justification = justified,singlelinecheck = false}
\resizebox{\columnwidth}{!}{%
\begin{tabular}{cccccccccc}
\hline
\multirow{2}{*}{}          & \multicolumn{3}{|c|}{\ToneTtwoPD} & \multicolumn{3}{|c|}{\ToneTtwoFlair} & \multicolumn{3}{|c|}{MRI $\rightarrow$ CT} \\ \cline{2-10} 
                           & PSNR      & SSIM & FID    & PSNR      & SSIM & FID     & PSNR      & SSIM & FID     \\ \hline
\multirow{2}{*}{ResViT}    &   \textbf{33.92}        &  \textbf{0.977}        &   \textbf{14.47}        &    \textbf{25.84}      & 0.886          &  \textbf{18.58}  &    28.45      & 0.931          &  60.28      \\
                           &    \textbf{$\pm$1.44 }      &  \textbf{$\pm$0.004}        &           &   \textbf{$\pm$1.13}       &   $\pm$0.014        &        &   $\pm$1.35       &   $\pm$0.009        &      \\ \hline

w/o trans.     &  32.91         &    0.966      & 14.56  &   24.96      &    0.868      &  19.21&  26.73        &    0.899   & 95.38  \\
            modules             &    $\pm$0.96       &  $\pm$0.005        & & $\pm$1.10         &   $\pm$0.005       &  & $\pm$0.91        &     $\pm$0.008      &       \\ \hline
                        
w/o conv.   &     33.49      &      0.971    &     14.84      &    25.11      &    0.874 &     20.30   &    28.19      &    0.922 &     \textbf{60.16}        \\
modules 
                           &    $\pm$1.34       &  $\pm$0.005        &           &   $\pm$1.02       &   $\pm$0.014        &      &   $\pm$1.15       &   $\pm$0.009  &              \\ \hline
                           
w/o adv.     &  33.75         &    \textbf{0.977}      &    15.80      &    22.95      &   \textbf{0.891}        &    40.68   &   \textbf{28.58}      &   \textbf{0.932}        &    65.49   \\
      loss                    &    $\pm$1.45      &  \textbf{$\pm$0.005}       &           &   $\pm$1.93       &   \textbf{$\pm$0.015}        & 
                        &    \textbf{ $\pm$1.13  }       &   \textbf{$\pm$0.007}       &                   \\ \hline
\end{tabular}
}
\caption{\revhl{Test performance of ResViT and variants ablated of transformer modules, convolutional modules or adversarial loss. FID is a summary metric across the entire test set. Boldface indicates the top-performing model for each task.}}
\label{tab:ablation_1}
\end{table}

\renewcommand{\tabcolsep}{6pt}
\renewcommand{\arraystretch}{1.20}
\begin{table}[h]
\centering
\captionsetup{justification = justified,singlelinecheck = false}
\resizebox{0.9\columnwidth}{!}{%
\begin{tabular}{ccccccc}
\hline
\multirow{2}{*}{}          & \multicolumn{2}{|c|}{\ToneTtwoPD} & \multicolumn{2}{|c|}{\ToneTtwoFlair} & \multicolumn{2}{|c|}{MRI $\rightarrow$ CT} \\ \cline{2-7} 
                           & PSNR      & SSIM     & PSNR      & SSIM      & PSNR      & SSIM     
                           \\ \hline
\multirow{2}{*}{$A_1 - A_6$}    &   \textbf{33.92}        &  \textbf{0.977}        &   \textbf{25.84}        &    \textbf{0.886}      & \textbf{28.45}          &  \textbf{0.931}      
\\
                       &    \textbf{$\pm$1.44 }      &  \textbf{$\pm$0.004}                  &   \textbf{$\pm$1.13}       &   \textbf{$\pm$0.014}             &   \textbf{$\pm$1.35}      &   \textbf{$\pm$0.009}              \\ \hline

\multirow{1}{*}{$A_1 - A_6$}      &  33.72         &    0.973      & 25.19  &   0.879      &    28.16      &  0.923 \\
              (untied weights)          &    $\pm$1.23       &  $\pm$0.005        &$\pm$1.18 & $\pm$ 0.014        &   $\pm$1.11    &   $\pm$0.007           \\ \hline
                        
\multirow{2}{*}{$A_1$}   &     33.51      &      0.971    &     24.98      &    0.883      &    28.06 &     0.921          \\

                           &    $\pm$1.15       &  $\pm$0.005        &     $\pm$1.60      &   $\pm$0.015       &   $\pm$1.31           &   $\pm$0.008               \\ \hline
                           
\multirow{2}{*}{$A_6$}      &  33.78         &    \textbf{0.977}      &    25.25      &    0.880      &   27.95        &    0.921      \\
                        &    $\pm$1.34       &  \textbf{$\pm$0.004}      &      $\pm$1.20     &   $\pm$0.014      &   $\pm$1.22     &   $\pm$0.008                   \\ \hline
\end{tabular}
}
\caption{\revhl{Test performance of ResViT {\small($A_1 - A_6$)} and variants ablated of weight tying and individual transformer modules. Boldface indicates the top-performing model for each task.}}
\label{tab:ablation_2}
\end{table}

\renewcommand{\tabcolsep}{6pt}
\renewcommand{\arraystretch}{1.20}
\begin{table}[h!]
\centering
\captionsetup{justification = justified,singlelinecheck = false}
\resizebox{0.9\columnwidth}{!}{%
\begin{tabular}{ccccccc}
\hline
\multirow{2}{*}{}          & \multicolumn{2}{|c|}{\ToneTtwoPD} & \multicolumn{2}{|c|}{\ToneTtwoFlair} & \multicolumn{2}{|c|}{MRI $\rightarrow$ CT} \\ \cline{2-7} 
                           & PSNR      & SSIM     & PSNR      & SSIM      & PSNR      & SSIM     
                           \\ \hline
\multirow{2}{*}{ResViT}    &   \textbf{33.92}        &  \textbf{0.977}        &   \textbf{25.84}        &    \textbf{0.886}      & \textbf{28.45}          &  \textbf{0.931}      
\\
                   &    \textbf{$\pm$1.44 }      &  \textbf{$\pm$0.004}                  &   \textbf{$\pm$1.13}       &   \textbf{$\pm$0.014}             &   \textbf{$\pm$1.35}      &   \textbf{$\pm$0.009}       \\ \hline

\multirow{2}{*}{w/o pre-training}      &  33.55         &    0.971      & 24.86  &   0.881      &    27.94      &  0.912 \\
                      &    $\pm$1.25       &  $\pm$0.005        &$\pm$1.28 & $\pm$ 0.016        &   $\pm$1.25    &   $\pm$0.009           \\ \hline
                        
w/o del.   &     33.35      &      \textbf{0.977}    &     24.89      &    0.873      &    28.01 &     0.924          \\

                   insertion    &    $\pm$1.13       &  \textbf{$\pm$0.004}        &     $\pm$1.18      &   $\pm$0.015       &   $\pm$1.27           &   $\pm$0.008               \\ \hline
                           
w/o pre-training   &  33.58         &    0.971      &    24.74      &    0.869      &   27.66        &    0.913      \\
              or del. insertion        &    $\pm$1.16       &  $\pm$0.005       &      $\pm$1.30     &   $\pm$0.016       &   $\pm$0.78      &   $\pm$0.006                   \\ \hline
\end{tabular}
}
\caption{\revhl{Test performance of ResViT and variants ablated of pre-training and delayed insertion procedures for transformers. Boldface indicates the top-performing model for each task.}}
\label{tab:ablation_3}
\vspace{-1mm}
\end{table}

\revhl{Third, we examined the utility of the skip connections and down/upsampling blocks in the proposed architecture. We compared ResViT against variants built by removing the skip connection around the transformer module or around the CNN module in ART blocks. Table \ref{tab:ablation_4} lists performance metrics in the test set. ResViT yields higher performance than all variants (p$<$0.05). Our results indicate that ResViT benefits substantially from residual learning in ART blocks. We also compared ResViT against variants built by replacing down/upsampling modules in ART blocks with unlearned maxpooling/bilinear interpolation modules, and by increasing encoder downsampling and decoder upsampling rates to remove down/upsampling modules in ART blocks entirely. ResViT outperforms all variants as listed in Table \ref{tab:ablation_4} (p$<$0.05), except for MRI$\rightarrow$CT where the variant with unlearned down/upsampling and ResViT yield similar SSIM. These results demonstrate the benefits of the proposed down/upsampling scheme in ResViT.}

\revhl{Next, we inspected the relative strength of transformer-derived contextual features in distilled representations within ART blocks. To do this, we computed the $L_2$-norms of contextual feature maps output by the transformer module, and input feature maps from the previous ART block relayed through the skip connection of the transformer module. We also computed the relative weighting of the two feature maps as the $L_2$-norms of respective combination weights in the channel compression (CC) module. Measurements for ResViT models trained in representative tasks are listed in Table \ref{tab:ablation_5}. We find that contextual and input feature maps, and their respective combination weights in CC blocks have comparable strength, demonstrating that contextual features are a substantial component of image representations in ART blocks.}

\renewcommand{\tabcolsep}{6pt}
\renewcommand{\arraystretch}{1.20}
\begin{table}[t]
\vspace{2mm}
\centering
\captionsetup{justification = justified,singlelinecheck = false}
\resizebox{\columnwidth}{!}{%
\begin{tabular}{ccccccc}
\hline
\multirow{2}{*}{}          & \multicolumn{2}{|c|}{\ToneTtwoPD} & \multicolumn{2}{|c|}{\ToneTtwoFlair} & \multicolumn{2}{|c|}{MRI $\rightarrow$ CT} \\ \cline{2-7} 
                           & PSNR      & SSIM     & PSNR      & SSIM      & PSNR      & SSIM     
                           \\ \hline
\multirow{2}{*}{ResViT}    &   \textbf{33.92}        &  \textbf{0.977}        &   \textbf{25.84}        &    \textbf{0.886}      & \textbf{28.45}          &  \textbf{0.931}      
\\
                   &    \textbf{$\pm$1.44 }      &  \textbf{$\pm$0.004}                  &   \textbf{$\pm$1.13}       &   \textbf{$\pm$0.014}             &   \textbf{$\pm$1.35}      &   \textbf{$\pm$0.009}       \\ \hline

                        
w/o skip around   &     28.24      &      0.942    &     25.02      &    0.864      &    26.94 &     0.906          \\

      conv. modules    &    $\pm$1.27       &  $\pm$0.009        &     $\pm$0.98      &   $\pm$0.016       &   $\pm$0.73           &   $\pm$0.007              \\ \hline
w/o skip around   &     31.53      &      0.962    &     24.06      &    0.868      &    27.08 &     0.908          \\

         trans. modules                  &    $\pm$1.26       &  $\pm$ 0.006        &     $\pm$1.28      &   $\pm$0.014       &   $\pm$0.80           &   $\pm$0.006               \\ \hline
ART with unlearned    &  33.73         &    0.969      &    25.33      &    0.884      &   28.16        &    \textbf{0.931}      \\
      down/upsampling       &    $\pm$1.19       &  $\pm$0.005       &      $\pm$1.11     &   $\pm$0.014       &   $\pm$1.04      &   \textbf{$\pm$0.007}                   \\ \hline
ART w/o   &  31.51         &    0.961      &    23.61      &    0.867      &   26.79        &    0.915      \\
     down/upsampling                 &    $\pm$1.27       &  $\pm$0.006       &      $\pm$1.53     &   $\pm$0.015       &   $\pm$0.62      &   $\pm$0.006                   \\ \hline
\end{tabular}
}
\caption{\revhl{Test performance of ResViT and variants built by: removing skip connections in convolutional modules, removing skip connections in transformer modules, using unlearned down/upsampling blocks in ART, removing down/upsampling blocks in ART via a higher degree of down/upsampling in the encoder/decoder. Boldface indicates the top-performing model for each task.}}
\label{tab:ablation_4}
\end{table}

\begin{table}[h!]
\centering
\captionsetup{justification = justified,singlelinecheck = false}
\resizebox{0.9\columnwidth}{!}{%
\begin{tabular}{cccc}
\hline
    & \multicolumn{1}{|c|}{\ToneTtwoPD} & \multicolumn{1}{|c|}{\ToneTtwoFlair} & \multicolumn{1}{|c|}{MRI $\rightarrow$ CT}   
                           \\ \hline
\multirow{1}{*}{$g$}   &     277.88      &      400.90    &     421.46            \\\hline

\multirow{1}{*}{$f$}      &  536.48        &    571.23 &    636.93      \\\hline

\multirow{1}{*}{CC weights for $g$}    &   96.5       &  112.04        &  72.88  
\\\hline

\multirow{1}{*}{CC weights for $f$}      &  226.08         &    169.15     & 116.70  \\\hline
\end{tabular}
}
\caption{\revhl{Feature maps and corresponding combination weights for the channel compression (CC) module were inspected in ResViT. Averaged across the test set and ART blocks, $L_2$-norm of feature maps from the transformer module ($g$) and feature maps input by the previous ART block ($f$) are listed along with combination weights for $g$ and for $f$.}}
\label{tab:ablation_5}
\vspace{-1mm}
\end{table}

Lastly, we wanted to visually interpret the benefits of the self-attention mechanisms in ResViT towards synthesis performance. Fig. \ref{fig:adv_attention_merge}b displays representative attention maps in ResViT. Synthetic images and error maps are also shown for ResViT as well as pGAN, which generally offered the closest performance to ResViT in our experiments. \revhl{We find that the attention maps exhibit higher intensity in critical regions such as brain lesions in multi-contrast MRI and pelvic bone structure in MR-to-CT synthesis. Importantly, these regions of higher attentional focus are also the primary regions where the synthesis errors are substantially diminished with ResViT compared to pGAN. Taken together, these results suggest that the transformer-based ResViT model captures contextual relationships related to both healthy and pathological tissues to improve synthesis performance.} 


\section{Discussion}
\par
In this study, we proposed a novel adversarial model for image translation between separate modalities. Traditional GANs employ convolutional operators that have limited ability to capture long-range relationships among distant regions \cite{kodali2018}. The proposed model aggregates convolutional and transformer branches within a residual bottleneck to preserve both local precision and contextual sensitivity. To our knowledge, this is the first adversarial model for medical image synthesis with a transformer-based generator. We further introduced a weight-sharing strategy among transformer modules to lower model complexity. Finally, a unification strategy was implemented to learn an aggregate model that copes with numerous source-target configurations without training separate models. 

\par
\revhl{We demonstrated ResViT for missing modality synthesis in multi-contrast MRI and MRI-CT imaging. ResViT outperformed several state-of-the-art convolutional and transformer models in one-to-one and many-to-one tasks. We trained all models with an identical loss function to focus on architectural influences to synthesis performance. In unreported experiments, we also trained competing methods that were proposed with different loss functions using their original losses, including PTNet with mean-squared loss \cite{ptnet} and medSynth with mean-squared, adversarial and gradient-difference losses \cite{nie2018}. We observed that ResViT still maintains similar performance benefits over competing methods in these experiments. Yet, it remains important future work to conduct an in-depth assessment of optimal loss terms for ResViT, including gradient-difference and difficulty-aware losses for the generator \cite{nie2018,Luping1,Shen1}, and edge-preservation and binary cross-entropy losses for the discriminator \cite{Shen2,Shen1}.}

\par
Trained with image-average loss terms, CNNs have difficulty in coping with atypical anatomy that substantially varies across subjects \cite{pix2pix,pgan}. To improve generalization, recent studies have proposed self-attention mechanisms in GAN models over spatial or channel dimensions \cite{attention_unet,sagan}. Specifically, attention maps are used for multiplicative modulation of CNN-derived feature maps. This modulation encourages the network to focus on critical image regions with relatively limited task performance. While attention maps can be distributed across image regions, they mainly capture implicit contextual information via modification of local CNN features. Since feature representations are primarily extracted via convolutional filtering, the resulting model can still manifest limited expressiveness for global context. In contrast, the proposed architecture uses dedicated transformer blocks to explicitly model long-range spatial interactions in medical images.  

\par
Few recent studies have independently proposed transformer-based models for medical image synthesis \cite{ganbert,ptnet,kamran2021}. In \cite{ganbert}, a transformer is included in the discriminator of a traditional GAN for MR-to-PET synthesis. In \cite{ptnet}, a UNet-inspired transformer architecture is proposed for infant MRI synthesis \cite{ptnet}. Differing from these efforts, our work makes the following contributions. (1) Compared to \cite{ganbert} that uses transformers to learn a prior for target PET images, we employ transformers in ResViT's generator to learn latent contextual representations of source images. (2) Unlike \cite{ptnet} that uses mean-squared error loss amenable to over-smoothing of target images \cite{pgan}, we leverage an adversarial loss to preserve realism. (3) While \cite{ptnet} uses a convolution-free transformer architecture, we instead propose a hybrid architecture that combines localization capabilities of CNNs with contextual sensitivity of transformers. (4) While \cite{ganbert} and \cite{ptnet} consider only task-specific, one-to-one synthesis models, here we uniquely introduce many-to-one synthesis models and a unified model that generalizes across multiple source-target configurations.

\par
\revhl{UNet-style models follow an encoder-decoder architecture with an hourglass structure \cite{pix2pix}. Because spatial resolution is substantially lower in the midpoint of the hourglass (e.g. 16x16 maps), these models typically introduce skip connections between the encoder and decoder layers to facilitate preservation of low-level features. In contrast, ResViT is a ResNet-style model where encoded representations pass through a bottleneck of residual blocks before reaching the decoder \cite{resnet}, and encoder-decoder skip connections are omitted due to several reasons. First, ResViT maintains relatively high resolution at the output of its encoder (e.g. 64x64 maps), so its bottleneck represents relatively lower-level information. Second, each ART block is organized as a transformer-CNN cascade with skip connections around both modules, creating a residual path between the input and output of each block. This eventually bridges the encoder output to the decoder input, creating a native residual path in ResViT. Lastly, we observed during early stages of the study that a variant model that included encoder-decoder skip connections caused a minor performance drop, suggesting that these extra connections might reduce the effectiveness of the central information bottleneck.}

\par
\revhl{Here, ResViT models were initialized with transformers pre-trained on 16x16 input feature maps. In turn, 256x256 images were 16-fold downsampled cumulatively across the encoder and transformer modules, and the transformer used a patch size of P=1 and sequence length of 256. Several strategies can be adopted to use ResViT at different image resolutions. In a first scenario, the downsampling rate and patch size can be preserved, while the sequence length is adjusted. For instance, a 512x512 image would be downsampled to a 32x32 feature map, resulting in a sequence of 1024 patches. While a transformer pre-trained on 32x32 maps would be ideal, vision transformers can reliably handle variable sequence lengths without retraining so the original transformer can still be used \cite{vit}. Note that longer sequences would incur a quadratic increase in processing and memory load in both cases \cite{vit}. In a second scenario, the original transformer with sequence length 256 can be maintained, while either the patch size or the downsampling rate is adjusted. For a 512x512 image, P=2 (2x2 patches) on a 32x32 map (16-fold downsampled) or P=1 on a 16x16 map (32-fold downsampled) could be used. Both options would achieve on par computational complexity to the original architecture, albeit the transformer would process feature maps at a relatively lower resolution compared to the resolution of the input image. It is unlikely that this would significantly affect ResViT’s sensitivity to local features since the primary component of ART that captures local features is the residual CNN module whose resolution can be preserved. If the input image does not have a power-of-two size, the abovementioned strategies can be adopted after zero-padding to round up the resolution to the nearest power of two, or by implementing the encoder with non-integer downsampling rates \cite{chen2021convolutional}. Note that computer vision studies routinely fine-tune transformers at different image resolutions than encountered during pre-training without performance loss \cite{vit}, so ResViT might also demonstrate similar behavior. It remains important future work to investigate the comparative utility of the discussed resolution-adaptation strategies in medical image synthesis.}

\par
Several lines of development can help further improve ResViT's performance. Here, we considered synthesis tasks in which source and target modalities were registered prior to training, and they were paired across subjects. When registration accuracy is limited, a spatial registration block can be incorporated into the network. Furthermore, a cycle-consistency loss \cite{cyclegan} can be incorporated in the optimization objective to allow the use of unregistered images. This latter strategy would also permit training of ResViT models on unpaired datasets \cite{woltering2017,ge2019}. Data requirements for model training can be further alleviated by adopting semi-supervised strategies that allow mixing of paired and unpaired training data \cite{jin2018}, or that would enable training of synthesis models directly from undersampled acquisitions \cite{yurt2021ss}. \revhl{Finally, ResViT might benefit from incorporation of multi-scale modules in the decoder to improve preservation of fine image details \cite{Shen2}}.

\vspace{-0.5ex}
\section{Conclusion}
Here we introduced a novel synthesis approach for multi-modal imaging based on a conditional deep adversarial network. In an information bottleneck, ResViT aggregates convolutional operators and vision transformers, thereby improving capture of contextual relations while maintaining localization power. A unified implementation was introduced that prevents the need to rebuild models for varying source-target configurations. ResViT achieves superior synthesis quality to state-of-the-art approaches in multi-contrast brain MRI and multi-modal pelvic MRI-CT datasets. Therefore, it holds promise as a powerful candidate for medical image synthesis.  

\vspace{-0.5ex}
\bibliographystyle{IEEETran} 
\bibliography{IEEEabrv,refs}

\end{document}